# How to manipulate magnetic states of antiferromagnets

Cheng Song[1,2], Yunfeng You[1], Xianzhe Chen[1,2], Xiaofeng Zhou[1,2], Yuyan Wang[3], Feng Pan[1,2]

[1] Key Laboratory of Advanced Materials (MOE), School of Materials Science and Engineering, Tsinghua University, Beijing 100084, China

[2] Beijing Innovation Center for Future Chip, Tsinghua University, Beijing 100084, China.

[3] Department of Physics, Beihang University, Beijing 100191, China

E-mail: songcheng@mail.tsinghua.edu.cn (C.S.)

Abstract:

Antiferromagnetic materials, which have drawn considerable attention recently, have fascinating features: they are robust against perturbation, produce no stray fields, and exhibit ultrafast dynamics. Discerning how to efficiently manipulate the magnetic state of an antiferromagnet is key to the development of antiferromagnetic spintronics. In this review, we introduce four main methods (magnetic, strain, electrical, and optical) to mediate the magnetic states and elaborate on intrinsic origins of different antiferromagnetic materials. Magnetic control includes a strong magnetic field, exchange bias, and field cooling, which are traditional and basic. Strain control involves the magnetic anisotropy effect or metamagnetic transition. Electrical control can be divided into two parts, electric field and electric current, both of which are convenient for practical applications. Optical control includes thermal and electronic excitation, an inertia-driven mechanism, and terahertz laser control, with the potential for ultrafast antiferromagnetic manipulation. This review sheds light on effective usage of antiferromagnets and provides a new perspective on antiferromagnetic spintronics.

## Introduction

Spintronics, also called spin electronics, studies the effective control and manipulation of spin degrees of freedom in solid-state systems [1]. During the past three decades, spintronics has been widely researched worldwide, and its potential usage has drawn significant attention. Conventional spintronic devices depend on the manipulation of magnetic moments in ferromagnets [2]. However, as antiferromagnets exhibit myriad intriguing features, devices whose key parts are made of antiferromagnets have garnered much interest. They have inspired a new field called antiferromagnetic spintronics with the aim of complementing or replacing ferromagnets in the active components of spintronic devices [3]. Compared to ferromagnets, antiferromagnetic materials resist perturbation well, produce no stray fields, demonstrate ultrafast dynamics, and generate large magneto-transport effects [1, 3-9]. Accordingly, antiferromagnetic materials have become increasingly important and exhibit various promising applications including non-volatile memory [2, 10, 11] and magnetic field probes [12].

In 2011, a spin-valve-like magnetoresistance of an antiferromagnet-based tunnel junction marked the formal emergence of antiferromagnetic spintronics [13]. Thereafter, extensive researches were conducted in this area, focused on the important roles antiferromagnets play in spintronics. Barely a year later, tunneling anisotropic magnetoresistance of antiferromagnetic tunnel junctions was realized at room temperature [14]. In 2014, a room-temperature antiferromagnetic memory resistor [11] as well as a strain- and electric field-controlled metamagnetic transition emerged [15], promoting the application of antiferromagnetic spintronics. 2016 witnessed the reversible electrical switching of antiferromagnets [6], paving the way for antiferromagnetic information storage devices. In fact, antiferromagnetic materials were studied much earlier, mainly focusing on exchange bias and the spin-flop field [5].

Recently, many new phenomena have arisen in antiferromagnets such as the large anomalous Hall effect [12, 16-18], spin Hall magnetoresistance [19-23], skyrmions [24-28], and the spin Seebeck effect [29-33], enriching the microcosmic physics system and encouraging the vigorous development of antiferromagnetic spintronics. With antiferromagnets, field-free switching of magnetization through spin–orbit torque has also been realized [34-37]. Therefore, determining how to manipulate the magnetic states of antiferromagnets efficiently is key to further development [5]. Research efforts have aimed to solve this problem via magnetic, strain, electrical, and optical manipulation. Magnetic control includes a strong magnetic field, exchange bias and field cooling, which is fundamental. Strain control involves the magnetic anisotropy effect or metamagnetic transition. Electrical control can be divided into two parts: electric field and electric current, which are convenient for practical applications. Optical control includes thermal and electronic excitation, an inertia-driven mechanism, and terahertz laser control, which has the potential for ultrafast antiferromagnetic manipulation.

Some reviews of antiferromagnetic spintronics over the last decade have explicated its intrinsic theoretics and introduced various experimental methods [3-5, 12, 38-42]. For instance, a comprehensive review in 2016 illustrated the physical roots and spintronic devices related to different types of antiferromagnetic materials [5]. Other reviews pertained to information reading, writing, and storage in antiferromagnetic materials [38]; elaborated spin-transfer torques, giant-magnetoresistance effects, and exchange bias in antiferromagnetic metal spintronics [4]; or discussed half-metallic antiferromagnetic spintronics in detail [39]. Nevertheless, to the best of our knowledge, no review has systematically examined diverse manipulation approaches of antiferromagnetic states. In this review, we focus on recent works that have adopted diverse methods (as shown in figure 1) to mediate antiferromagnets' magnetic states and elaborate on their unique intrinsic origins, thereby illuminating the effective usage of antiferromagnets and providing a new perspective on antiferromagnetic spintronics.

**Magnetic control of antiferromagnets**

Magnetic field-related methods offer common and effective ways to manipulate the direction of magnetic moments, especially for ferromagnets. For antiferromagnets, however, it is much harder because of the vanishing applicable Zeeman energy [5]. Recent research has shown that the magnetic moment of antiferromagnets can also be controlled similarly to ferromagnets [3, 38]. On the one hand, the magnetic moments can be manipulated merely by applying an external magnetic field that exceeds a certain value, but the magnetic field required for aligning magnetic moments or domains is quite strong. In NiO(111) single crystals, the required magnetic field is 9 T, aligning the magnetic moments perpendicular to the magnetic field direction due to Zeeman energy reduction aside from contributions of magnetic anisotropy or domain formation via magnetostriction [43]. For $Mn_2Au$, 70 T in-plane magnetic field is imperative to realizing an antiferromagnetic domain orientation [44].

Exchange bias, on the other hand, is more universal for controlling the magnetic moments of antiferromagnets attached to ferromagnets with a smaller magnetic field. Exchange bias was first discovered by Meiklejohn and Bean in 1956 [45], after which it was widely researched for over half a century. We can observe exchange bias in the magnetic hysteresis loop as its center shifts along the horizontal axis, which arises from the interfacial coupling of ferromagnets and antiferromagnets when heterostructures are field cooled below their respective Curie ($T_C$) and Néel ($T_N$) temperatures [46, 47], as presented in figure 2.

Considering interface coupling mechanisms, different kinds of exchange bias are suitable for various materials [46]. Parallel ferromagnet/antiferromagnet coupling is explained in systems such as Co/$LaFeO_3$/$SrTiO_3$(001) [48], CoFe/NiO(111) [49], Co/NiO(001) [50], MnPd(001)/Fe/MgO(001) [51], FeMn/Co [52], and IrMn(002)/Fe/MgO(001) [53]. Antiparallel coupling has been discovered in materials like Co/$FeF_2$/$MgF_2$(110) [54, 55] and CoO/CoPt multilayers [56]. Perpendicular coupling (spin-flop) is found in multi-layers (e.g., CoO/Fe/Ag(001) [57], Co/NiO(001) [58], NiO/Fe(001) [59], Fe/NiO/Ag(001) [60], $Fe_3O_4$/CoO(001) [61], and IrMn/[Co/Pt] [62]). In addition, out-of-plane coupling has also been identified in epitaxial films like FeMn/Ni/Cu(001) [63]. Additionally, with diverse thicknesses and measurement temperatures, the same system could show different coupling modes [46]. Actually, any factor that influences the coupling of ferromagnets and antiferromagnets could contribute directly to changes in exchange bias [46, 64]. For instance, film thickness [13, 63, 65-68], temperature [65, 69-71], atomic-steps [72, 73], substrates [60], interlayer spacer [74-76], and layer termination [77, 78] can change either the antiferromagnetic anisotropy or modifications of the interfacial characteristics. In addition, when uncompensated interfacial antiferromagnetic spins couple to the external field, extrinsic parameters such as field cooling may also affect exchange bias greatly [46]. A positive exchange bias can be obtained at lower cooling magnetic fields, while higher cooling magnetic fields are required to obtain negative exchange bias because of the competing coupling between antiferromagnet-ferromagnet and antiferromagnet-external fields [46, 64, 79, 80] or competition between the antiferromagnetic coupling at the interface and the Zeeman energy of the antiferromagnetic spins [81]. The microstructure of different samples largely determines the magnitude of the cooling field required to obtain a positive or negative exchange bias [47]. Antiferromagnetic interface spin arrangements or compensation is crucial for exchange bias as well. Compensated versus uncompensated antiferromagnetic surfaces [55, 82-85] and in-plane versus out-of-plane antiferromagnetic spins [63, 86-88] are central to this issue, which is closely related to antiferromagnetic layer anisotropy and microstructure (such as grain size and roughness) in distinct crystalline orientations [47]. Large anisotropy could result in bulk effects for antiferromagnets, causing interfacial spins to remain in their bulk configuration [89-91] while varied microstructures decided by specific growth conditions could affect the interface states [47].

Although the exchange bias of ferromagnet/antiferromagnet bilayers or multilayers has been studied for several decades, studies have mainly emphasized ferromagnets and the interface. Antiferromagnets are only passively used to pin ferromagnets [1, 4, 5, 46]. Not until recently have antiferromagnets shown promising applications in spintronics and come to play an increasingly active role in exchange bias, through which the magnetic moments in antiferromagnets can be manipulated effectively [5, 46]. Magnetoresistance phenomena provide methods for studying the exchange bias and magnetic characteristics of antiferromagnetic films. Considering the aforementioned concepts, multi-layer stacks have been researched extensively, with antiferromagnets playing a leading role. When applying an external magnetic field, the moments of ferromagnetic layer rotate with it directly. Then, moments of antiferromagnetic layer tilt because of exchange bias, forming an exchange spring to transfer the modulation of magnetic moments from ferromagnets to antiferromagnets [92, 93]. The exchange bias (exchange spring) of magnetized ferromagnets and antiferromagnets can be demonstrated via positive/negative field-cooled magnetization loops [3, 5, 13, 46].

Exchange spring is a vital assumption of models describing exchange bias, having been proven experimentally in ferromagnets [94, 95] and antiferromagnets [96]. For an antiferromagnet, if magnetic moments of the ferromagnet beside it are switched or rotated, a planar domain wall is wound up in the antiferromagnet [97], forming a spring-like structure due to exchange bias [96]. Magnetic control of antiferromagnets using the mechanism above was initially realized in the NiFe/IrMn/MgO/Pt system, where an IrMn antiferromagnet was laid next to the tunnel barrier, controlling the transport signal as shown in figure 3. The change of exchange-spring action and the rotation of antiferromagnets' magnetic moments is reflected in antiferromagnetic tunnelling anisotropic magnetoresistance (TAMR) and superconducting quantum interference device measurements [98]. In addition, the influence of temperature, amplitude of the field that triggers magnetic moment rotation, and thickness of antiferromagnetic films are all essential for altering magnetic moments in antiferromagnets [13].

Note that the large TAMR obtained by Park *et al.* is in a low-temperature condition (4 K, far less than room temperature), limiting the practical usage for information storage. Later, TAMR was found to be significantly enhanced at room temperature, achieved via perpendicular magnetic anisotropy (PMA) magnetic tunnel junctions [14]. PMA magnetic tunnel junctions are promising devices for new-generation nonvolatile memory ascribed to the reduction of the critical current in spin-transfer torque switching [99, 100]. Giant perpendicular exchange bias has already been found in multilayers [101]. To exceed the temperature limitations of TAMR application in antiferromagnets, [Pt/Co]/IrMn-based junctions are first utilized to obtain a room temperature TAMR effect with the exchange bias mechanism. As shown in figure 4, a squared vertical hysteresis loop presents a strong PMA of the junctions when applying vertical magnetic fields (the direction of the easy axis). For parallel magnetic fields, the hysteresis loop becomes a line. PMA of the films lays the foundation for the exchange bias of ferromagnets and antiferromagnets, being the premise of room temperature TAMR due to superior thermal tolerance and stable antiferromagnetic moments in relatively thick IrMn. The TAMR signal resembles the hysteresis loop, exhibiting a hysteresis window with a stable high-resistance state at a primitively applied positive $H$ as well as a low-resistance state at negative $H$ afterwards. The TAMR signal originates from the partial rotation of antiferromagnetic IrMn moments induced by Co/Pt magnetization. The high-resistance state arises from the perpendicular relationship of Co/Pt moments and IrMn moments when a vertical positive $H$ is applied because the Co/Pt moments rotate along their easy axis while IrMn moments are intrinsically aligned in-plane. A low-resistance state otherwise originates from the quasi-parallel relationship for negative $H$, ascribed to the occurrence of incomplete rotation of IrMn moments with a rotation angle, forming an exchange spring coupled to Co/Pt as shown in the insets [14].

Similar anisotropic magnetotransport phenomena such as anisotropic magnetoresistance (AMR) were discovered later in other systems like semiconductor $Sr_2IrO_4$ [102, 103], insulator $Y_3Fe_5O_{12}$ [104], and FeRh [11, 105]. At the same time, manipulating magnetic moments in antiferromagnets through exchange bias has also been demonstrated in materials such as $Fe_2CrSi/Ru_2MnGe$ epitaxial bilayers [106],

Fe/CuMnAs bilayers [85], [Co/Pt]/FeMn [107], and $La_{0.67}Sr_{0.33}MnO_3$/G-$SrMnO_3$/$La_{0.67}Sr_{0.33}MnO_3$ (LSMO/SMO/LSMO) sandwiches [108].

However, using exchange bias to manipulate the magnetic moments in an antiferromagnet is limited by the thickness of the antiferromagnetic layer because the exchange spring required to trigger the rotation of antiferromagnetic moments can only occur when the thickness does not exceed the domain wall width of the antiferromagnet [13, 14]. Therefore, the thickness should be within a narrow range, maintaining a careful balance between thinness (for exchange-spring rotation to cross the width of the antiferromagnet) and thickness (for size effects to avoid the descending Néel temperature, which is quite strict for practical applications). Besides, exchange bias-based devices can be interfered with by weak magnetic field perturbations because of the ferromagnets within them. To solve this problem, magnetic control of antiferromagnets without ferromagnetic elements has emerged in recent years, opening a new path for antiferromagnetic memory resistors using field cooling. Different resistance states can be designed by field cooling the devices from above their Néel temperature along different orientations [5]. For Ta/MgO/IrMn tunneling junctions, this process occurs due to the formation of different antiferromagnetic configurations of uniform IrMn film. The field cooling procedure may favor spin configurations with different proportions of distinct metastable antiferromagnetic phases, causing a shift in the metastable resistance [109]. For the FeRh memory resistor of FeRh/MgO structure shown in figure 5, the mechanism behind the field cooling effect is different, as there is an antiferromagnetic–ferromagnetic transition of about 400 K [11]. FeRh has ferromagnetic orders at high temperatures, convenient to set distinct collective directions of Fe moments by applying the necessary magnetic field. When the resistor is field cooled to room temperature, antiferromagnetic orders appear, and the direction of the antiferromagnetic moments is predetermined by the magnetic field and moment direction in a high-temperature ferromagnetic state. Different resistances produced by magnetic moment reorientation are essential to memory resistors; this manipulation of antiferromagnets facilitates ultrafast, precession-free writing schemes [11].

To sum up, magnetic control of antiferromagnets is an important issue involving various ways of manipulating the magnetic moments in antiferromagnets. Of them, manipulation via exchange bias is basic and universal, accompanied by other methods such as strong magnetic field or field cooling. Different magnetic manipulation methods of antiferromagnets are summarized in table 1; however, magnetic control of antiferromagnets warrants continued research and development to further diminish the applied magnetic field as well as to increase resistance to external perturbations.

## Strain control of antiferromagnets

To avoid perturbations in the external magnetic field and to control magnetic moments in antiferromagnets without the influence of ferromagnets, strain is proposed as an intriguing means of manipulation via either the magnetic anisotropy effect or metamagnetic transition. For the former, strain exerts an influence on the lattice parameter and spin-orbit coupling, which has a strong relationship with magnetic anisotropies related to the ground-state energy, chemical potential, and density of states, being the origin of antiferromagnets' magnetic moment orientation [110]. It is worth noting that this method has been predicted and observed primarily in bimetallic antiferromagnets such as $Mn_2Au$ [44, 110] and MnIr [62], due to large spin-orbit coupling on the 5*d* shell of the noble metal as well as the large moment on the Mn 3*d* shell. To illustrate, we take $Mn_2Au$ to demonstrate the effects of strain on magnetic moments. The staggered moments of $Mn_2Au$ could rotate in different directions by applying various sufficiently strong strain along the easy axis triggered by magnetocrystalline anisotropy energy; the exact directional relationship between strain and rotation has been extracted [110]. For example, strain may cause lattice distortion and reduce in-plane symmetry from 4-fold to 2-fold in $Mn_2Au$(001) crystal structure, leading to an uniaxial anisotropy with preference for a certain Néel vector orientation compared to samples without strain. Therefore, the magnetic moments rotate spontaneously. Moreover, Mn spins in $Mn_2Au$ prioritize the alignment along the direction with the shorter lattice spacing [44].

The other mechanism relies on the metamagnetic transition from the ferromagnetic to antiferromagnetic phase [111]. Recent works in this field have focused mostly on strained FeRh thin films [112, 113]. The metamagnetic transition temperature from antiferromagnetic order to ferromagnetic order is approximately 350 K, meaning that FeRh may exhibit an antiferromagnetic state at room temperature [114-119], which is suitable for information storage [120]. A lattice expansion of 1% is noticeable, accompanied by the phase transition from antiferromagnet to ferromagnet in which point strain produced during the epitaxial growth process can also control the metamagnetic transition by modifying the electronic structure under tetragonal lattice distortion, resulting in magnetocrystalline anisotropy in thin films. Therefore, different substrates beneath the films might exert influence on the strain in FeRh [113, 121]. For instance, FeRh grown on thick single-crystalline MgO and ion-beam-assist-deposited (IBAD) MgO exhibits metamagnetic transition, but the Fe moments show different orientations in FeRh. Furthermore, IBAD MgO can better match the FeRh ferromagnetic phase, thus lowering the transition temperature by stabilizing this phase spontaneously [121]. In addition, the experimental conditions also influence the strain in the epitaxial films, which deserves close attention.

With the rapid development of antiferromagnetic spintronics, a low-power method is required to efficiently control antiferromagnet magnetism [2]. Use of the strain-related metamagnetic transition of FeRh triggered by the electric field to realize low-power manipulation is promising for magnetic storage. In this method, $BaTiO_3$ is needed to dominate the metamagnetic transition temperature of FeRh films because of its ferroelectricity [122]; thus, the transition between antiferromagnetic and ferromagnetic order happens just above room temperature with only a few volts. This electric-field control process can be summarized by two points: piezoelectricity provides voltage-induced strain, and the field effect depletes or accumulates carriers in a material adjacent to the ferroelectric part. The pioneering work of Cherifi *et al.* indicated that a low electric field, such as $E$ = 0.4 kV cm$^{-1}$, can cause the metamagnetic transition temperature to increase by about 25 K, which is adequate to isothermally convert the FeRh from the antiferromagnetic to ferromagnetic order. If the electric field is turned off, FeRh could restore the initial antiferromagnetic state. As we mentioned above, the underlying cause for this phenomenon lies in electric field-induced strain, which can alter the lattice parameter and electronic structure of FeRh. First-principles calculations demonstrate that the antiferromagnetic state is more stable without strain and with an increase in compressive strain. $\Delta E$ (the energy difference between ferromagnet $E_{\mathrm{FM}}$ and antiferromagnet $E_{\mathrm{AFM}}$, which can be expressed as: $\Delta E = E_{\mathrm{FM}} - E_{\mathrm{AFM}}$.) is displayed in figure 6; it increases monotonically, suggesting that the antiferromagnetic phase becomes progressively stable [15].

The theoretical calculation aligns with the experimental results, providing verification that the strain can alter the magnetic order in antiferromagnet FeRh [15].

The metamagnetic transition for the FeRh/$BaTiO_3$ structure is further explained by the transition origins from mutual mechanisms of the polarization-reversal-induced volume/strain expansion in FeRh at the interface and the competition between magnetic exchange couplings [123]. The unique properties of the metamagnetic transition could play an important part in antiferromagnet strain control, benefiting the development of strain-controlled antiferromagnets.

In this section, we have discussed the influence of strain on the manipulation of antiferromagnets and introduced several ways to control antiferromagnetic moments related to strain. Different strain-related manipulation methods of antiferromagnets are summarized in table 2. Compared to the magnetic control of antiferromagnets, the range for strain manipulation is narrower and often combined with other effects such as the electric field. However, this topic still inspires international research interest given its unique characteristics such as energy saving and ease of reaching room temperature [15]. Note that manipulation through strain has been reported and achieved in metallic antiferromagnets, mainly in $Mn_2Au$, MnIr, and FeRh. There have been few reports about oxides and other antiferromagnetic materials. The essence of strain-controlled antiferromagnets lies in the change of electronic structure and strain-induced spin-orbit coupling, which can be applied to other materials. To this point, it is possible to control magnetic moments of more antiferromagnetic materials through strain or strain-related methods, which may expand the application of antiferromagnetic spintronics.

## Electrical control of antiferromagnets

As mentioned in the previous section, strain combined with an electric field can control the magnetic order of antiferromagnets above room temperature. In fact, the electric field can manipulate antiferromagnetic moments in itself or in combination with other methods [124], constituting an important part of antiferromagnetic spintronics [125]. Coupling between magnetization and the electric field by multiferroic, magnetoelectric materials or exchange bias helps to realize the manipulation of antiferromagnets and the design of low-power spintronics architectures such as information storage devices. Relying on the combination of field cooling and large magnetic fields or subsidiary ferromagnets to alter the magnetic configuration is not so convenient. Hence, manipulation by electric current has become more popular in recent years as an innovative and effective method that is not attached to ferromagnets or complicated procedures. In this section, we focus on the various electrical methods of antiferromagnetic manipulation, highlighting manipulation mechanisms for the electric field and current alike.

Early work regarding the manipulation of antiferromagnets in the electric field was conducted using multiferroic films. In this case, two or more ferroic orders coexist in multiferroic materials; for example, a large ferroelectric polarization and a small magnetization are observed in $BiFeO_3$ thin films with a large magnetoelectric coupling [126]. Ferroelectric polarization is switchable by electric field [127], and different polar states correspond to different signs of the chirality of the antiferromagnetic domains, meaning that antiferromagnetic domains can be controlled by electric field [128-130]. Among cycloidal (helicoidal) magnets, $BiFeO_3$ appears promising because the control process can occur at room temperature rather than at a very low temperature [125]. The key to electrical control of antiferromagnetic domains in multiferroic $BiFeO_3$ films at room temperature lies in the coupling between ferroelectricity and antiferromagnetism in $BiFeO_3$ thin films. More specifically, the switching, originating from the coupling of antiferromagnetic and ferroelectric domains to the underlying ferroelastic domain structure, has been demonstrated experimentally by piezoelectric force microscopy (PFM), photoemission electron microscopy (PEEM) and theoretically by first-principles calculation, as shown in figure 7 [131].

For $BiFeO_3$ single crystals, however, antiferromagnetic domains can be controlled by an electric field through strain-induced redistribution of the occupations of equivalent magnetic domains in addition to reversible ferroelectric domain switching [132]. In addition, coupling between the antiferromagnetic and ferroelectric orders in $BiFeO_3$ may be stronger in the bulk than in thin films where the cycloid is absent [133]. Similar strain-induced switching of antiferromagnetic moments modulated by ferroelectric substrates has been discovered in Ni/NiO heterostructures [134]. Recently, the magnetotransport and electronic transport in $BiFeO_3$ were found to occur across domain walls by external fields [135, 136], clarifying the manipulation mechanism of $BiFeO_3$ antiferromagnetic moments and promoting the development of multiferroic materials in spintronics.

Apart from multiferroic materials, magnetoelectric $Cr_2O_3$ is also used to electrically control antiferromagnetic domains; this is vitally important, as global magnetization reversal and reversible isothermal magnetoelectric switching can be realized at room temperature. In (0001) surface of antiferromagnetic $Cr_2O_3$, two degenerate 180° antiferromagnetic domains are controlled by magnetoelectric annealing. After magnetoelectric annealing, a single-domain antiferromagnetic state emerges with all surface Cr spins pointing in the same direction, and the spin-polarized surface is insensitive to surface roughness [137]. Switchable exchange bias can be realized when the electrically switchable $Cr_2O_3$ is coupled with ferromagnets [137-140]. Purely antiferromagnetic magnetoelectric random access memory using $Cr_2O_3$ as the antiferromagnetic element has also been designed, opening an appealing avenue for magnetoelectric antiferromagnet research [141].

Regarding metallic antiferromagnets such as IrMn and FeMn, the above mechanism no longer applies; metallic antiferromagnets do not have ferroelectric orders or magnetoelectric properties. However, metallic antiferromagnets are irreplaceable in traditional spintronics, so direct electrical control of them is significant yet challenging. The difficulty lies in the screening effect by the surface charge, confining the manipulation to a limited depth of atomic dimensions, which is not deep enough to form a stable antiferromagnetic exchange spring. To solve this problem, ionic liquid is used as the dielectric gate to regulate the exchange spring in antiferromagnets, as shown in figure 8 [142].

When dropping ionic liquid onto the surface of structures that are under control, an electric double layer (EDL) forms over the channel surface to accumulate electron carriers, exhibiting an extremely high electric-field effect and penetrating more deeply in antiferromagnets compared to conventional solid gate insulators [142-146]. The exchange spring formed in antiferromagnets can be controlled by the electric field. The exchange spring, as we discussed in the previous section, can alter the reorientation of antiferromagnetic moments. This method has been successfully applied in the [Co/Pt]/IrMn system, where an ionic liquid electrolyte allows the electric field to control the magnetic moments by using a different gate voltage ($V_G$). Negative $V_G$ has been found to strengthen the negative exchange bias while positive $V_G$ has the opposite effect, suggesting that $V_G$ can modify the exchange spring's stability. The intrinsic physics of this method is that the change of charge carriers in IrMn realized by injecting or extracting electrons from the film through the shift of $V_G$ could alter the electronic structure and magnetic moment of Mn, thus regulating the magnetic anisotropy and exchange bias that are used to observe the change of magnetic moments in antiferromagnets. However, the manipulation of IrMn by the electric field becomes weaker with increasing thickness of IrMn and disappears if the thickness exceeds the domain wall width (about 6 nm) [142]. A similar phenomenon is also discovered in FeMn, with a longer depth of domain wall width (about 15 nm) due to the larger magnetic anisotropy of the top FeMn layer [147]. The progress in the electrical control of antiferromagnets through an ionic liquid provides a novel method for manipulation combining the electric field and exchange spring [148].

To manipulate the antiferromagnetic moments directly without ferromagnets, magnetic fields, or heating, an electric current becomes the priority with ultrafast spin dynamics [149, 150]. Spin-orbit torque plays an important role in this mechanism to achieve all-electrical isothermal manipulation of antiferromagnetic moments at room temperature [125]. When electric currents flow through the solid, the relativistic current-induced spin-orbit coupling interaction can stimulate a magnetic field on every atomic site. However, a non-zero effect on a particular lattice site cannot appear unless the crystal structure has a broken inversion symmetry, by which means the sign of the generated magnetic field may change with the reversal of the electrical current direction as well as the reflection of the central atom’s atomic environment. In this point, as long as the chemical environment around the magnetic moments is alternated, then an alternate magnetic torque would emerge, acting on each magnetic moment. Therefore, manipulating antiferromagnetic moments by electrical currents requires a special crystal structure, matching the magnetic modulation by chemical modulation [6, 125, 151]. Antiferromagnetic $Mn_2Au$ is first predicted to have the required crystal structure suitable for switching via electric currents. The lattice of $Mn_2Au$ can be divided into two sublattices, each possessing broken inversion symmetry and giving opposite inverse spin galvanic effects. At the same time, the two sublattices can form inversion partners, which allows for the observation of Néel-order spin-orbit torque (NSOT) fields [125, 151, 152]. The first experimental proof of NSOT switching by electric current is in CuMnAs, which has a similar crystal structure to $Mn_2Au$ with two sublattices symmetric with respect to Mn sites, as illustrated in figure 9(a). An alternated magnetic field is produced in the whole thin film as the electric current flows along the basal plane. Then, the staggered Mn magnetic moments rotate simultaneously following spatially modulated torque, finally remaining stable along particular directions controlled by magnetic anisotropies. To see the macroscopic change of switching, a CuMnAs compact device is designed with a cross-bar structure, allowing for sequential writing and reading as presented in figure 9(b), (c). When electrical pulses are applied to the device along [100] or [010] directions, the magnetic moments of Mn rotate to be perpendicular to the current. The electrical resistance of the CuMnAs bar alters due to anisotropic magnetoresistance. The direct observation of current-induced switching of antiferromagnetic domains in CuMnAs is made possible through X-ray magnetic linear dichroism microscopy, which shows a clear correlation between the average domain orientation and the anisotropy of electrical resistance [6]. In addition, the Néel vector in a CuMnAs thin-film can also be determined by optical measurements [153]. Using this type of method, stable and switchable memory states can be obtained; if shorter pulses are applied, the CuMnAs devices can incrementally transit from two opposite saturating resistance states [6, 125, 154].

Similar works were soon reported about $Mn_2Au$ [155, 156], a good conductor with a high Néel temperature (above 1000 K), which is suitable for memory applications. Moreover, $Mn_2Au$ is a good metallic conductor without toxic components and has better thermal stability for applications than CuMnAs. The reliable and reproducible switching of antiferromagnetic moments is actualized by current pulses while the read-out is conducted by magnetoresistance measurements and the planar Hall effect. Besides the commitment of the electric current, thermal activation is also highly important in the reorientation process of the magnetization states of $Mn_2Au$, proving that electric control of magnetization can be stable at room temperature for a long time, thus boosting the development of this material in memory devices [6, 155, 156].

Above all, electrical control of antiferromagnets is prosperous for application in storage devices because the manipulation process can be conducted at room temperature, with no need for a magnetic field, field cooling, or ferromagnets. Different electrical manipulation methods of antiferromagnets are summarized in table 3. Nevertheless, materials that are suitable for electric manipulation are usually confined to either multiferroic material such as $BiFeO_3$, magnetoelectric material like $Cr_2O_3$, metallic antiferromagnets that can form an exchange spring such as IrMn, or antiferromagnets with globally noncentrosymmetric unit cells such as $Mn_2Au$. Therefore, more materials suitable for electrical control of their magnetic moments should be explored and utilized to enhance the range of materials considered appropriate for practical usage. Recently, current-induced switching is also realized in ferrimagnets with high thermal stability, such as Pt/CoGd [157] and Ta/$Co_{1-x}Tb_x$ system [158], as well as synthetic antiferromagnets [159, 160]. Besides, the electric current density required for switching is $4.5\times10^6$ A $cm^{-2}$ for CuMnAs [6] and $2\times10^7$ A $cm^{-2}$ for $Mn_2Au$ [156], which should be further decreased to create energy-saving devices.

## Optical control of antiferromagnets

Apart from the methods mentioned above, optical manipulation is another feasible way to realize effective control of antiferromagnets. Similar to ferromagnets, lights can be used to control an antiferromagnet's magnetization via optically driven thermal and electronic excitation, as demonstrated by second-harmonic generation, linear reflection, and other advanced technologies. The interaction of light and magnetism paves the way for light to probe or control magnetic materials with an ultrafast timescale, which has attracted considerable attention in fundamental physics and technological applications of magnetic recording and information processing [161-164]. The ultrafast dynamics of antiferromagnetic spins via optical excitation has been realized in several kinds of antiferromagnets through different mechanisms. For $Cr_2O_3$, ultrafast spin dynamics are due to the spin–lattice interaction and emission of magnetic excitations during nonradiative carrier decay [163, 165-167]. In terms of NiO, ultrafast antiferromagnetic reorientation of $Ni^{2+}$ spins stems from the ultrafast transient antiferromagnetic phase transition between hard- and easy-axis states via a string of pump pulses that change the magnetic anisotropy. Second-harmonic generation is used to probe the change [163, 168-170]. As for rare-earth orthoferrites ($RFeO_3$; R indicates a rare-earth element) such as $TmFeO_3$, when a short femtosecond laser pulse is absorbed in the materials, a rapid temperature-dependent anisotropy change may be induced and it can lead to the reorientation of the spins through spin–lattice interaction [7, 171, 172]. Another type of important material whose magnetic states can be controlled by optical excitations is the colossal-magnetoresistance compound $Pr_{1-x}Ca_xMnO_3$ or $Pr_{1-x}Na_xMnO_3$, where an ultrafast transition from an insulating antiferromagnet to a conducting metallic state can be triggered by laser [163, 173-176]. Similarly, through a phase transition, the spin configuration of antiferromagnet $DyFeO_3$ would be changed into a noncollinear spin state with a net magnetization from the collinear compensated state [177].

In addition to these ultrafast dynamics, a fundamentally different scenario of spin switching in antiferromagnets has been discovered that has a slow response to an optical stimulus. Inertial behavior is exhibited due to the exchange interaction between spins, the process of which is as follows and is depicted in figure 10(a): a strong magnetic-field pulse generated by laser transfers sufficient momentum and kinetic energy to the spin system despite that spin orientation hardly changes during the action; however, the momentum is strong enough that spins overcome the potential barrier and reorient into a new metastable state long after the pulse, thus realizing switching in antiferromagnets. Inertia-driven spin switching has been observed in antiferromagnets like $HoFeO_3$ [178, 179] and $YFeO_3$ [180], providing new insight into ultrafast recording and information storage.

We can employ new developments in terahertz technology to control the spin degree of freedom in antiferromagnets apart from manipulating charge motion. Terahertz radiation is located between infrared light and microwave radiation, the frequency of which is approximately 0.1 to 10 THz, providing a generic ultrafast method to dominate previously inaccessible magnetic excitations in the electronic ground state. Earlier studies show that low-energy elementary excitations in solids like lattice vibrations and plasma oscillations can be manipulated by the terahertz field, which controls charge carriers through electric-dipole coupling. Magnetic terahertz interaction with magnetic-dipole coupling is much harder to harness due to the weak magnetic dipoles; even so, it is possible to realize ultrafast spin control in antiferromagnets because the spin of an electron is associated with a magnetic moment, and the time-dependent magnetic field exerts a Zeeman torque that can be used to control the spin [181-183]. High-intensity terahertz pulses have been applied to investigate the spin reorientation, specifically the interplay between antiferromagnets and the magnetic component of intense terahertz transients, as shown in figure 10(b) [182].

The magnetic component couples to electron spins selectively decided via Zeeman interaction. The Zeeman torque enables the spins to turn out of the easy plane, and the anisotropy field forces them into precession motion with respect to their equilibrium direction. The out-of-plane spin component is precisely antiparallel in the two sublattices, whereas the in-plane coordinate oscillates in the same direction. Note that in terahertz systems, whose observed dynamics are solely motivated by the magnetic rather than the electric field, the terahertz pulse is aimed exclusively at the spins and does not deposit excessive heat in other degrees of freedom, unlike the mechanism of optical pulses heating the electron system. For example, the temperature of the excited NiO volume by terahertz pump pulse increases less than 10 μK even with the effects of complete relaxation of the magnon wave, thus qualifying the ultrafast coherent control of spins [182]. Besides using the magnetic field component of terahertz pulses to coherently control magnons in the electronic ground state by direct Zeeman interaction, nonlinearity of the terahertz electric field can also be used to control antiferromagnetic orders [181], which is much stronger than linear Zeeman coupling to the terahertz magnetic field. This is because the coupling of electronic orbital states to ordered spins in nearly all materials is highly correlated with the strength and direction of the magnetic anisotropy, at which point an ultrafast electric field pulse induced by intense terahertz pulses can change the orbital state of electrons abruptly and suddenly alter the magnetic anisotropy. Finally, nonlinear spin control is realized by triggering coherent magnon oscillations. This mechanism has successfully been demonstrated in antiferromagnetic $TmFeO_3$ [184]. Antiferromagnetic $HoFeO_3$ [185] and NiO [186] also exhibit nonlinear magnetization dynamics but are still related to the terahertz magnetic field. Therefore, there are two complementary concepts of fully coherent coupling of nonlinearities depending on terahertz pulses: the magnetic field and electric field component of terahertz pulses [187]. The observation and control of the spin waves and resonance [188] in the time domain via terahertz technology have expanded to other antiferromagnetic materials such as MnO, $NdFeO_3$, $YFeO_3$, $TbMnO_3$, $HoMnO_3$, and so on. The spin wave of MnO is excited only by linearly polarized laser pulses ascribed to angular momentum conservation [189], whereas for $NdFeO_3$ and $YFeO_3$, the spin wave is driven by temperature-dependent in-plane anisotropy [190] and dielectric anisotropy of the crystal [191-193], respectively. As for $TbMnO_3$ [194], $HoMnO_3$ [193], and $YMnO_3$ [196], the spin dynamics is driven by a terahertz antiferromagnetic resonance. These varied means of effective ultrafast spin control of antiferromagnets renders it possible to develop future information technologies, such as ultrafast data recording and quantum computation. To achieve practical applications of opto-magnetism, reversible switching of antiferromagnetism is essential and has been demonstrated in a helical antiferromagnet $TbMnO_3$ as of late, using a two-color-pump laser set-up. Optical control is achieved by an electric polarization induced by antiferromagnetic order; reversible switching is then realized via a two-color approach, where the direction of the antiferromagnetic order parameter reverses with the changing of the incident light wavelength. Besides, the optically induced antiferromagnetic switching remains stable over time and is under local control, allowing for the generation and deletion of high-density antiferromagnetic domain walls, so associated magnetoresistance can emerge in the process; see figure 10(c) [162]. The reversible switching of antiferromagnetic domains provides a way to store information using an optical method.

In this section, different optical methods are discussed that control the antiferromagnetic spins with an emphasis on spin dynamics in antiferromagnets with the effects of femtosecond laser pulses. Examples involve light-induced spin reorientation in a canted antiferromagnet, photo-induced antiferromagnet–ferromagnet phase transitions, and driving spin precessions via magneto–phononic coupling. Spin waves and resonant coupling induced by intense terahertz pulses are also mentioned, presenting linear and nonlinear magnetization dynamics through different mechanisms. Moreover, the magnetic field and electric field of terahertz pulses are complementary for ultrafast magnetization control. Different optical manipulation methods of antiferromagnets are summarized in table 4. Optical methods to control antiferromagnetism uncover ultrafast information recording and processing for purely optical memory devices in the future.

## Conclusion

In summary, the purpose of this review is to walk the readers through recent works on the manipulation of antiferromagnets with the aim of connecting past achievements with the promise of ongoing developments in the field of antiferromagnetic spintronics. Although more studies have begun addressing the manipulation of antiferromagnetic states via diverse methods, the field is still in its nascent stage. It requires extensive research to investigate the instinct mechanism of different kinds of antiferromagnetic materials and the relationship between external fields and antiferromagnets to devise energy-saving and simple methods for practical applications. In this review, we have summarized four main methods (magnetic, strain, electrical, and optical) to control antiferromagnets; these methods represent most of the findings in this field to date. Magnetic manipulation is traditionally used in ferromagnetic spintronics, especially with regard to exchange bias, as the pinning layer to modulate ferromagnetic moments. However, with the rapid development of antiferromagnetic spintronics, exchange bias has become increasingly important to manipulate the magnetic structure of antiferromagnets by itself or in combination with other fields. On the other hand, the formation of exchange bias needs ferromagnets, indicating that it is easy to be interfered by external perturbation. Besides, compared with ultrafast switching of spins via laser, the velocity of exchange bias-controlled spin reorientation is much slower. Therefore, other methods are proposed to solve these problems.

Through strain-induced magnetic anisotropy or metamagnetic transition, the spin configuration can be changed without ferromagnets at room temperature. Electrical control of antiferromagnets can realize swift switching as well, require no magnetic field, field cooling, or ferromagnets. These excellent characteristics show promise for future spintronics devices like memory resistors. In contrast, the strain and electric methods are confined to a small range of materials and demand a distinctive crystal structure. As for the optical means, spin dynamics are being extensively studied, providing ultrafast spin manipulation on a femtosecond scale. They show potential for the design of ultrafast information processing and recording devices. Nevertheless, the mechanism for optical methods is complex, and there is presently little connection between theory and application. The disadvantages and alluring potential of these methods compel researchers to make progress in this area. Knowing that methods of controlling antiferromagnets exceed what we have mentioned, it would be impossible for us to cover all relevant aspects in this paper. Other unique methods for specific materials are intriguing as well, complementing the entire system of manipulation.

To stimulate continued advancement of the field, more suitable and functional materials should be investigated along with novel methods that are efficient in energy and operation. The inner mechanism of interaction between spin and different methods also deserves close attention, providing guidance for experimental research. Above all, effective manipulation of magnetic states in antiferromagnets is central to the application and further development of antiferromagnetic spintronics.

**Acknowledgements**

C.S. acknowledges the support of Young Chang Jiang Scholars Program. This work was financially supported by the National Natural Science Foundation of China (Grant Nos. 51571128, 51671110 and 51601006).

**Reference**

[1] Žutić I, Fabian J and Sarma S D 2004 Spintronics: fundamentals and applications *Rev. Mod. Phys.* **76** 323-410

[2] Chappert C, Fert A and Van Dau F N 2007 The emergence of spin electronics in data storage *Nature Mater.* **6** 813-23

[3] Gomonay O, Jungwirth T and Sinova J 2017 Concepts of antiferromagnetic spintronics *Phys. Status Solidi-R* **11** 1700022

[4] Macdonald A H and Tsoi M 2011 Antiferromagnetic metal spintronics *Phil. Trans. R. Soc. A* **369** 3098-114

[5] Jungwirth T, Marti X, Wadley P and Wunderlich J 2016 Antiferromagnetic spintronics *Nature Nanotech.* **11** 231-41

[6] Wadley P *et al* 2016 Electrical switching of an antiferromagnet *Science* **351** 587-90

[7] Kimel A V, Kirilyuk A, Tsvetkov A, Pisarev R V and Rasing T 2004 Laser-induced ultrafast spin reorientation in the antiferromagnet $TmFeO_3$ *Nature* **429** 850-3

[8] Zhang W, Jungfleisch M B, Jiang W, Pearson J E, Hoffmann A, Freimuth F and Mokrousov Y 2014 Spin Hall effects in metallic antiferromagnets *Phys. Rev. Lett.* **113** 196602

[9] Shiino T, Oh S H, Haney P M, Lee S W, Go G, Park B G, and Lee K J 2016. Antiferromagnetic domain wall motion driven by spin-orbit torques *Phys. Rev. Lett.* **117** 087203

[10] Marrows C 2016 Addressing an antiferromagnetic memory *Science* **351** 558-9

[11] Marti X *et al* 2014 Room-temperature antiferromagnetic memory resistor *Nature Mater.* **13** 367-74

[12] Nakatsuji S, Kiyohara N and Higo T 2015 Large anomalous Hall effect in a non-collinear antiferromagnet at room temperature *Nature* **527** 212-5

[13] Park B G *et al* 2011 A spin-valve-like magnetoresistance of an antiferromagnet-based tunnel junction *Nature Mater.* **10** 347-51

[14] Wang Y Y, Song C, Cui B, Wang G Y, Zeng F and Pan F 2012 Room-temperature perpendicular exchange coupling and tunneling anisotropic magnetoresistance in an antiferromagnet-based tunnel junction *Phys. Rev. Lett.* **109** 137201

[15] Cherifi R O *et al* 2014 Electric-field control of magnetic order above room temperature *Nature Mater.* **13** 345-351

[16] Nayak A K *et al* 2016 Large anomalous Hall effect driven by a nonvanishing Berry curvature in the noncolinear antiferromagnet $Mn_3Ge$ *Sci. Adv.* **2** e1501870

[17] Kiyohara N, Tomita T and Nakatsuji S 2015 Giant anomalous Hall effect in the chiral antiferromagnet $Mn_3Ge$ *Phys. Rev. Applied* **5** 064009

[18] Sürgers C, Wolf T, Adelmann P, Kittler W, Fischer G and Löhneysen H V 2017 Switching of a large anomalous Hall effect between metamagnetic phases of a non-collinear antiferromagnet *Sci. Rep.* **7** 42982

[19] Zhou X, Ma L, Shi Z, Fan W J, Zheng J G, Evans R F L and Zhou S M 2015 Magnetotransport in metal/insulating-ferromagnet heterostructures: Spin Hall magnetoresistance or magnetic proximity effect *Phys. Rev. B* **92** 060402

[20] Han J H, Song C, Li F, Wang Y Y, Wang G Y, Yang Q H and Pan F 2014 Antiferromagnet-controlled spin current transport in $SrMnO_3$/Pt hybrids *Phys. Rev. B* **90** 144431

[21] Shang T *et al* 2016 Effect of NiO inserted layer on spin-Hall magnetoresistance in Pt/NiO/YIG heterostructures *Appl. Phys. Lett.* **109** 032410

[22] Lin W and Chien C L 2017 Electrical detection of spin backflow from an antiferromagnetic insulator/$Y_3Fe_5O_{12}$ interface *Phys. Rev. Lett* **118** 067202

[23] Hou D, Qiu Z, Barker J, Sato K, Yamamoto K, Vélez S, Gomez-Perez J M, Hueso L E, Casanova F and Saitoh E 2017 Tunable sign change of spin Hall magnetoresistance in Pt/NiO/YIG structures *Phys. Rev. Lett* **118** 147202

[24] Barker J and Tretiakov O A 2016 Static and dynamical properties of antiferromagnetic skyrmions in the presence of applied current and temperature *Phys. Rev. Lett.* **116** 147203

[25] Raičević I, Popović D, Panagopoulos C, Benfatto L, Silva Neto M B, Choi E S and Sasagawa T 2011 Skyrmions in a doped antiferromagnet. *Phys. Rev. Lett.* **106** 227206

[26] Buhl P M, Freimuth F, Blügel S and Mokrousov Y 2017 Topological spin Hall

effect in antiferromagnetic skyrmions *Phys. Status Solidi-R* **11** 1700007

[27] Göbel B, Mook A, Henk J and Mertig I 2017 Antiferromagnetic skyrmion crystals: generation, topological Hall, and topological spin Hall effect *Phys. Rev. B* **96** 060406

[28] Zhang X, Zhou Y and Ezawa M 2015 Antiferromagnetic skyrmion: stability, creation and manipulation *Sci. Rep.* **6** 24795

[29] Wu S M, Zhang W, Kc A, Borisov P, Pearson J E, Jiang J S, Lederman D, Hoffmann A and Bhattacharya A 2016 Antiferromagnetic spin Seebeck effect *Phys. Rev. Lett.* **116** 097204

[30] Seki S, Ideue T, Kubota M, Kozuka Y, Takagi R, Nakamura M, Kaneko Y, Kawasaki M and Tokura Y 2015 Thermal generation of spin current in an antiferromagnet *Phys. Rev. Lett.* **115** 266601

[31] Mendes J B S, Cunha R O, Alves Santos O, Ribeiro P R T, Machado F L A, Rodríguez-Suárez R L, Azevedo A and Rezende S M 2014 Large inverse spin Hall effect in the antiferromagnetic metal $Ir_{20}Mn_{80}$ *Phys. Rev. B* **89** 140406

[32] Rezende S M, Rodríguezsuárez R L and Azevedo A 2016 Theory of the spin Seebeck effect in antiferromagnets *Phys. Rev. B* **93** 014425

[33] Prakash A, Brangham J, Yang F and Heremans J P 2016 Spin Seebeck effect through antiferromagnetic NiO *Phys. Rev. B* **94** 014427

[34] Oh, Y W *et al* 2016 Field-free switching of perpendicular magnetization through spin-orbit torque in antiferromagnet/ferromagnet/oxide structures *Nature Nanotech.* **11** 878-84

[35] Fukami S, Zhang C, DuttaGupta S, Kurenkov A and Ohno H 2016 Magnetization switching by spin-orbit torque in an antiferromagnet-ferromagnet bilayer system *Nature Mater.* **15** 535-41

[36] Lau Y C, Betto D, Rode K, Coey J M D and Stamenov P 2016 Spin-orbit torque switching without an external field using interlayer exchange coupling *Nature Nanotech.* **11** 758-62

[37] Yu G *et al* 2014 Switching of perpendicular magnetization by spin-orbit torques in the absence of external magnetic fields *Nature Nanotech* **9** 548-54

[38] Marti X, Fina I and Jungwirth T 2015 Prospect for antiferromagnetic spintronics *IEEE Trans. Magn.* **51** 1-4

[39] Hu X 2011 Half-metallic antiferromagnet as a prospective material for spintronics *Adv. Mater.* **24** 294-8

[40] Gomonay E V and Loktev V M 2014 Spintronics of antiferromagnetic systems *Low Temp. Phys.* **40** 17-35

[41] Wang Y Y, Song C, Zhang J Y and Pan F 2017 Spintronic materials and devices based on antiferromagnetic metals *Prog. Nat. Sci.-Mater.* **27** 208-16

[42] Coileáin C Ó and Wu H C 2017 Materials, devices and spin transfer torque in antiferromagnetic spintronics: a concise review *Spin* **7** 1740014

[43] Hoogeboom G R, Aqeel A, Kuschel T, Palstra T T M and Wees B J V 2017 Negative spin Hall magnetoresistance of Pt on the bulk easy-plane antiferromagnet NiO *Appl. Phys. Lett.* **111** 052409

[44] Sapozhnik A A, Abrudan R, Skourski Y, Jourdan M, Zabel H, Kläui M and Elmers H J 2017 Manipulation of antiferromagnetic domain distribution in $Mn_2Au$ by ultrahigh magnetic fields and by strain *Phys. Status Solidi-R* **11** 1600438

[45] Meiklejohn W H and Bean C P 1956 New magnetic anisotropy *IEEE Trans. Magn.* **102** 3866-76

[46] Zhang W and Krishnan K M 2016 Epitaxial exchange-bias systems: from

fundamentals to future spin-orbitronics *Mat. Sci. Eng. R* **105** 1-20

[47] Nogués J and Schuller I K 1999 Exchange bias *J. Magn. Magn. Mater.* **192** 203-32

[48] Nolting F *et al* 2000 Direct observation of the alignment of ferromagnetic spins by antiferromagnetic spins *Nature* **405** 767-9

[49] Zhu W, Seve L, Sears R, Sinkovic B and Parkin S S P 2001 Field cooling induced changes in the antiferromagnetic structure of NiO films *Phys. Rev. Lett.* **86** 5389-92

[50] Ohldag H, Scholl A, Nolting F, Anders S, Hillebrecht F U and Stöhr J 2001 Spin reorientation at the antiferromagnetic NiO(001) surface in response to an adjacent ferromagnet *Phys. Rev. Lett.* **86** 2878-81

[51] Blomqvist P, Krishnan K M and Ohldag H 2005 Direct imaging of asymmetric magnetization reversal in exchange-biased Fe/MnPd bilayers by X-ray photoemission electron microscopy *Phys. Rev. Lett* **94** 107203

[52] Bali R, Nelsoncheeseman B B, Scholl A, Arenholz E, Suzuki Y and Blamire M G 2009 Competing magnetic anisotropies in an antiferromagnet-ferromagnet-antiferromagnet trilayer *J. Appl. Phys.* **106** 277-80

[53] Zhang W, Bowden M E and Krishnan K M 2011 Competing effects of magnetocrystalline anisotropy and exchange bias in epitaxial Fe/IrMn bilayers *Appl. Phys. Lett.* **98** 092503

[54] Fitzsimmons M R, Kirby B J, Roy S, Li Z P, Roshchin I V, Sinha S K and Schuller I K 2007 Pinned magnetization in the antiferromagnet and ferromagnet of an exchange bias system *Phys. Rev. B* **75** 214412

[55] Roy S *et al* 2005 Depth profile of uncompensated spins in an exchange bias system *Phys. Rev. Lett.* **95** 047201

[56] Kappenberger P, Martin S, Pellmont Y, Hug H J, Kortright J B, Hellwig O and Fullerton E E 2003 Direct imaging and determination of the uncompensated spin density in exchange-biased CoO/(CoPt) multilayers *Phys. Rev. Lett* **91** 267202

[57] Wu J, Park J S, Kim W, Arenholz E, Liberati M, Scholl A, Wu Y Z, Hwang C and Qiu Z Q 2010 Direct Measurement of Rotatable and Frozen CoO Spins in Exchange Bias System of CoO/Fe/Ag(001) *Phys. Rev. Lett* **104** 217204

[58] Van D L G, Telling N D, Potenza A, Dhesi S S and Arenholz E 2011 Anisotropic x-ray magnetic linear dichroism and spectromicroscopy of interfacial Co/NiO(001) *Phys. Rev. B* **83** 064409

[59] Brambilla A, Portalupi M, Finazzi M, Ghiringhelli G, Duò L, Parmigiani F, Zacchigna M, Zangrando M and Ciccacci F 2004 Magnetic anisotropy of NiO epitaxial thin films on Fe(001) *J. Magn. Magn. Mater.* **272** 1221-2

[60] Kim W, Jin E, Wu J, Park J, Arenholz E, Scholl A, Hwang C and Qiu Z Q 2010 Effect of NiO spin orientation on the magnetic anisotropy of the Fe film in epitaxially grown Fe/NiO/Ag(001) and Fe/NiO/MgO(001) *Phys. Rev. B* **81** 174416

[61] Ijiri Y, Schulthess T C, Borchers J A, van der Zaag P J and Erwin R W 2007 Link between perpendicular coupling and exchange biasing in $Fe_3O_4$/CoO multilayers *Phys. Rev. Lett.* **99** 147201

[62] Wang Y Y, Song C, Zhang J Y and Pan F 2017 Role of an ultrathin platinum seed layer in antiferromagnet-based perpendicular exchange coupling and its electrical manipulation *J. Magn. Magn. Mater.* **428** 431-6

[63] Wu J, Choi J, Scholl A, Doran A, Arenholz E, Hwang C and Qiu Z Q 2009 Ni spin switching induced by magnetic frustration in FeMn/Ni/Cu(001) *Phys. Rev.*

*B* **79** 212411

[64] Shi Z, Du J and Zhou S M 2014 Exchange bias in ferromagnet/antiferromagnet bilayers *Chin. Phys. B* **23** 027503

[65] Reichlova H *et al* 2016 Temperature and thickness dependence of tunneling anisotropic magnetoresistance in exchange-biased Py/IrMn/MgO/Ta stacks *Mater. Res. Express* **3** 076406

[66] Wu J, Choi J, Scholl A, Doran A, Arenholz E, Wu Y Z, Won C, Hwang C and Qiu Z Q 2009 Element-specific study of the anomalous magnetic interlayer coupling across NiO spacer layer in Co/NiO/Fe/Ag(001) using XMCD and XMLD *Phys. Rev. B* **80** 012409

[67] Meng Y, Li J, Tan A, Jin E, Son J, Park J S, Doran A, Young A T, Scholl A and Arenholz E 2011 Element-specific study of epitaxial NiO/Ag/CoO/Fe films grown on vicinal Ag(001) using photoemission electron microscopy *Appl. Phys. Lett.* **98** 212508

[68] Wang B Y, Jih N Y, Lin W C, Chuang C H, Hsu P J, Peng C W, Yeh Y C, Chan Y L, Wei D H and Chiang W C 2011 Driving magnetization perpendicular by antiferromagnetic-ferromagnetic exchange coupling *Phys. Rev. B* **83** 104417

[69] Zhang W and Krishnan K M 2012 Spin-flop coupling and rearrangement of bulk antiferromagnetic spins in epitaxial exchange-biased Fe/MnPd/Fe/IrMn multilayers *Phys. Rev. B* **86** 054415

[70] Cao W N, Li J, Chen G, Zhu J, Hu C R and Wu Y Z 2011 Temperature-dependent magnetic anisotropies in epitaxial Fe/CoO/MgO(001) system studied by the planar Hall effect *Appl. Phys. Lett.* **98** 262506

[71] Suzuki I, Hamasaki Y, Itoh M and Taniyama T 2014 Controllable exchange bias in Fe/metamagnetic FeRh bilayers *Appl. Phys. Lett.* **105** 172401

[72] Kuch W, Offi F, Chelaru L I, Kotsugi M, Fukumoto K and Kirschner J 2002 Magnetic interface coupling in single-crystalline Co/FeMn bilayers *Phys. Rev. B* **65** 140408

[73] Li J, Przybylski M, Yildiz F, Fu X L and Wu Y Z 2011 In-plane spin reorientation transition in Fe/NiO bilayers on Ag(1,1,10) *Phys. Rev. B* **83** 3965-70

[74] Gruyters M and Schmitz D 2008 Microscopic nature of ferro- and antiferromagnetic interface coupling of uncompensated magnetic moments in exchange bias systems *Phys. Rev. Lett* **100** 077205

[75] Meng Y *et al* 2012 Magnetic interlayer coupling between antiferromagnetic CoO and ferromagnetic Fe across a Ag spacer layer in epitaxially grown CoO/Ag/Fe/Ag(001) *Phys. Rev. B* **85** 014425

[76] Normile P S, Toro J A D, Muñoz T, González J A, Andrés J P, Muñiz P, Galindo R E and Riveiro J M 2007 Influence of spacer layer morphology on the exchange-bias properties of reactively sputtered Co/Ag multilayers *Phys. Rev. B* **76** 104430

[77] Wang B Y, Chuang C H, Wong S S, Chiou J J, Lin W C, Chan Y L, Wei D H and Lin M T 2012 Flipping magnetization induced by noncollinear ferromagnetic-antiferromagnetic exchange coupling *Phys. Rev. B* **85** 094412

[78] Chen G, Li J, Liu F Z, Zhu J, He Y, Wu J, Qiu Z Q and Wu Y Z 2010 Four-fold magnetic anisotropy induced by the antiferromagnetic order in FeMn/Co/Cu(001) system *J. Appl. Phys.* **108** 073905

[79] Nogués J, Lederman D, Moran T J and Schuller I K 1996 Positive exchange bias in $FeF_2$-Fe bilayers *Phys. Rev. Lett* **76** 4624-7

[80] Leighton C, Nogués J, Jönsson-Åkerman B J and Schuller I K 2000 Coercivity

enhancement in exchange biased systems driven by interfacial magnetic frustration *Phys. Rev. Lett.* **84** 3466-9

[81] Yang D Z, Du J, Sun L, Wu X S, Zhang X X and Zhou S M 2005 Positive exchange biasing in GdFe/NiCoO bilayers with antiferromagnetic coupling *Phys. Rev. B* **71** 144417

[82] Blamire M and Hickey B 2006 Magnetic materials: compensating for bias *Nature Mater.* **5** 87-8

[83] Brück S, Schütz G, Goering E, Ji X and Krishnan K M 2008 Uncompensated moments in the MnPd/Fe exchange bias system *Phys. Rev. Lett.* **101** 126402

[84] Nayak A K *et al* 2015 Design of compensated ferrimagnetic Heusler alloys for giant tunable exchange bias *Nature Mater.* **14** 679-84

[85] Wadley P et al. 2017 Control of antiferromagnetic spin axis orientation in bilayer Fe/CuMnAs films *Sci. Rep.* **7** 11147

[86] Moran T J and Schuller I K 1996 Effects of cooling field strength on exchange anisotropy at permalloy/CoO interfaces *J. Appl. Phys.* **79** 5109-11

[87] Jungblut R, Coehoorn R, Johnson M T, Sauer Ch, van der Zaag P J, Ball A R, Rijks Th G S M, aan de Stegge J and Reinders A 1995 Exchange biasing in MBE-grown $Ni_{80}Fe_{20}/Fe_{50}Mn_{50}$ bilayers *J. Magn. Magn. Mater.* **148** 300-6

[88] Park C M, Min K I and Shin K H 1996 Effects of surface topology and texture on exchange anisotropy in NiFe/Cu/NiFe/FeMn spin valves *J. Appl. Phys.* **79** 6228-30

[89] Kim K Y, Kim J W, Choi H C, Teichert A, You C Y, Park S, Shin S C and Lee J S 2011 Long-range interlayer-coupled magnetization reversal mediated by the antiferromagnetic layer in Py/FeMn/CoFe trilayers *Phys. Rev. B* **84** 144410

[90] Morales R, Li Z P, Olamit J, Liu K, Alameda J M and Schuller I K 2009 Role of

the antiferromagnetic bulk spin structure on exchange bias *Phys. Rev. Lett.* **102** 097201

[91] Nam D N H, Chen W, West K G, Kirkwood D M, Lu J and Wolf S A 2008 Propagation of exchange bias in CoFe/FeMn/CoFe trilayers *Appl. Phys. Lett.* **93** 152504

[92] Gao T R, Yang D Z, Zhou S M, Chantrell R, Asselin P, Du J and Wu X S 2007 Hysteretic behavior of angular dependence of exchange bias in FeNi/FeMn bilayers *Phys. Rev. Lett.* **99** 057201

[93] Qiu X P, Yang D Z, Zhou S M, Chantrell R, O'Grady K, Nowak U, Du J, Bai X J and Sun L 2008 Rotation of the pinning direction in the exchange bias training effect in polycrystalline NiFe/FeMn bilayers *Phys. Rev. Lett.* **101** 147207

[94] Kneller E F and Hawig R 2003 The exchange-spring magnet: a new material principle for permanent magnets *IEEE Trans. Magn.* **27** 3560-88

[95] Morales R, Basaran A C, Villegas J E, Navas D, Soriano N, Mora B, Redondo C, Batlle X and Schuller I K 2015 Exchange-bias phenomenon: the role of the ferromagnetic spin structure *Phys. Rev. Lett.* **114** 097202

[96] Scholl A, Liberati M, Arenholz E, Ohldag H and Stöhr J 2004 Creation of an antiferromagnetic exchange spring *Phys. Rev. Lett.* **92** 247201

[97] Li J, Tan A, Ma S, Yang R F, Arenholz E, Hwang C and Qiu Z Q 2014 Chirality switching and winding or unwinding of the antiferromagnetic NiO domain walls in Fe/NiO/Fe/CoO/Ag (001) *Phys. Rev. Lett.* **113** 147207

[98] Martí X *et al* 2012 Electrical measurement of antiferromagnetic moments in exchange-coupled IrMn/NiFe stacks *Phys. Rev. Lett.* **108** 017201

[99] Mangin S, Ravelosona D, Katine J A, Carey M J, Terris B D and Fullerton E E 2006 Current-induced magnetization reversal in nanopillars with perpendicular

anisotropy *Nature Mater.* **5** 210-5

[100] Ikeda S, Miura K, Yamamoto H, Mizunuma K, Gan H D, Endo M, Kanai S, Hayakawa J, Matsukura F and Ohno H 2010 A perpendicular-anisotropy CoFeB-MgO magnetic tunnel junction *Nature Mater.* **9** 721-4

[101] Feng J, Liu H F, Wei H X, Zhang X G, Ren Y, Li X, Wang Y, Wang J P and Han X F 2017 Giant perpendicular exchange bias in a subnanometer inverted $(Co/Pt)_n$/Co/IrMn Structure *Phys. Rev. Applied* **7** 054005

[102] Fina I *et al* 2014 Anisotropic magnetoresistance in an antiferromagnetic semiconductor *Nature Commun.* **5** 4671

[103] Wang C, Seinige H, Cao G, Zhou J S, Goodenough J B and Tsoi M 2014 Anisotropic magnetoresistance in antiferromagnetic $Sr_2IrO_4$ *Phys. Rev. X* **4** 041034

[104] Zhou X, Ma L, Shi Z, Fan W J, Evans R F, Zheng J G, Chantrell R W, Mangin S, Zhang H W and Zhou S M 2015 Mapping motion of antiferromagnetic interfacial uncompensated magnetic moment in exchange-biased bilayers *Sci. Rep.* **5** 09183

[105] Moriyama T, Matsuzaki N, Kim K J, Suzuki I, Taniyama T and Ono T 2015 Sequential write-read operations in FeRh antiferromagnetic memory *Appl. Phys. Lett.* **107** 122403

[106] Hajiri T, Matsushita M, Ni Y Z and Asano H 2017 Impact of anisotropy on antiferromagnet rotation in Heusler-type ferromagnet/antiferromagnet epitaxial bilayers *Phys. Rev. B* **95** 134413

[107] Wang Y Y, Song C, Wang G Y, Zeng F and Pan F 2014 Evidence for asymmetric rotation of spins in antiferromagnetic exchange-spring *New J. Phys.* **16** 123032

[108] Mao H J, Li F, Xiao L R, Wang Y Y, Cui B, Peng J J, Pan F and Song C 2015 Oscillatory exchange bias effect in $La_{0.67}Sr_{0.33}MnO_3$/*G*-$SrMnO_3$/$La_{0.67}Sr_{0.33}MnO_3$ sandwiches *J. Phys. D: Appl. Phys.* **48** 295003

[109] Petti D *et al* 2013 Storing magnetic information in IrMn/MgO/Ta tunnel junctions via field-cooling *Appl. Phys. Lett.* **102** 192404

[110] Shick A B, Khmelevskyi S, Mryasov O N, Wunderlich J and Jungwirth T 2010 Spin-orbit coupling induced anisotropy effects in bimetallic antiferromagnets: A route towards antiferromagnetic spintronics *Phys. Rev. B* **81** 212409

[111] Heeger A J 1970 Pressure dependence of the FeRh first-order phase transition *J. Appl. Phys.* **41** 4751-2

[112] Stamm C *et al* 2008 Antiferromagnetic-ferromagnetic phase transition in FeRh probed by x-ray magnetic circular dichroism *Phys. Rev. B* **77** 184401

[113] Gray A X *et al* 2012 Electronic structure changes across the metamagnetic transition in FeRh via hard X-ray photoemission *Phys. Rev. Lett.* **108** 257208

[114] Jiang M, Chen X Z, Zhou X J, Wang Y Y, Pan F and Song C 2016 Influence of film composition on the transition temperature of FeRh films *J Cryst. Growth* **438** 19-24

[115] Mankovsky S, Polesya S, Chadova K, Ebert H, Staunton J B, Gruenbaum T, Schoen M A W, Back C H, Chen X Z and Song C 2017 Temperature-dependent transport properties of FeRh *Phys. Rev. B* **95** 155139

[116] Jiang M, Chen X Z, Zhou X J, Cui B, Yan Y N, Wu H Q, Pan F and Song C 2016 Electrochemical control of the phase transition of ultrathin FeRh films *Appl. Phys. Lett.* **108** 202404

[117] Chen X Z, Liu H, Yin L F, Chen H, Song C and Pan F 2017 Hall detection of

anisotropic domain walls during magnetic phase transition *J. Phys. D: Appl. Phys.* **50** 505004

[118] Chen X Z, Feng J F, Wang Z C, Zhang J, Zhong X Y, Song C, Jin L, Zhang B, Li F and Jiang M 2017 Tunneling anisotropic magnetoresistance driven by magnetic phase transition *Nature Commun.* **8** 449

[119] Lee Y *et al* 2015 Large resistivity modulation in mixed-phase metallic systems *Nature Commun.* **6** 5959

[120] Thiele J U, Maat S and Fullerton E E 2003 FeRh/FePt exchange spring films for thermally assisted magnetic recording media *Appl. Phys. Lett.* **82** 2859-61

[121] Bordel C, Juraszek J, Cooke D W, Baldasseroni C, Mankovsky S, Minár J, Ebert H, Moyerman S, Fullerton E E and Hellman F 2012 Fe spin reorientation across the metamagnetic transition in strained FeRh thin films *Phys. Rev. Lett.* **109** 117201

[122] Valencia S *et al* 2011 Interface-induced room-temperature multiferroicity in $BaTiO_3$ *Nature Mater.* **10** 753-8

[123] Odkhuu D 2017 Electric control of magnetization reorientation in FeRh/$BaTiO_3$ mediated by a magnetic phase transition *Phys. Rev. B* **96** 134402

[124] Song C, Cui B, Peng J J, Mao H J and Pan F 2016 Electrical control of magnetism in oxides *Chin. Phys. B* **25** 29-44

[125] Fina I and Marti X 2017 Electric Control of Antiferromagnets *IEEE Trans. Magn.* **53** 1-7

[126] Sando D, Barthélémy A and Bibes M 2014 $BiFeO_3$ epitaxial thin films and devices: past, present and future *J. Phys.: Condens. Matter* **26** 473201

[127] Eerenstein W, Mathur N D and Scott J F 2006 Multiferroic and Magnetoelectric Materials *Nature* **442** 759-65

[128] Fina I, Fabrega L, Marti X, Sánchez F and Fontcuberta J 2010 Magnetic switch of polarization in epitaxial orthorhombic $YMnO_3$ thin films *Appl. Phys. Lett.* **97** 232905

[129] Fina I, Fàbrega L, Martí X, Sánchez F and Fontcuberta J 2011 Chiral domains in cycloidal multiferroic thin films: switching and memory effects *Phys. Rev. Lett.* **107** 257601

[130] Fina I, Skumryev V, O'Flynn D, Balakrishnan G and Fontcuberta J 2013 Phase coexistence and magnetically tuneable polarization in cycloidal multiferroics *Phys. Rev. B* **88** 100403

[131] Zhao T *et al* 2006 Electrical control of antiferromagnetic domains in multiferroic $BiFeO_3$ films at room temperature *Nature Mater.* **5** 823-9

[132] Lee S, Ratcliff W, Cheong S W and Kiryukhin V 2008 Electric field control of the magnetic state in $BiFeO_3$ single crystals *Appl. Phys. Lett.* **92** 192906

[133] Lebeugle D, Colson D, Forget A, Viret M, Bataille A M and Gukasov A A 2008 Electric-field-induced spin flop in $BiFeO_3$ single crystals at room temperature *Phys. Rev. Lett.* **100** 227602

[134] Zhang Y J, Chen J H, Li L L, Ma J, Nan C W and Lin Y H 2017 Ferroelectric strain modulation of antiferromagnetic moments in Ni/NiO ferromagnet/antiferromagnet heterostructures *Phys. Rev. B* **95** 174420

[135] He Q *et al* 2012 Magnetotransport at domain walls in $BiFeO_3$ *Phys. Rev. Lett.* **108** 067203

[136] Lee J H, Fina I, Marti X, Kim Y H, Hesse D and Alexe M 2014 Spintronic functionality of $BiFeO_3$ domain walls *Adv. Mater.* **26** 7078-82

[137] He X, Wang Y, Wu N, Caruso A N, Vescovo E, Belashchenko K D, Dowben P A and Binek C 2010 Robust isothermal electric control of exchange bias at

room temperature *Nature Mater.* **9** 579-585

[138] Echtenkamp W, Street M, Mahmood A and Binek C 2017 Tuning the effective anisotropy in a voltage-susceptible exchange-bias heterosystem *Phys. Rev. Applied* **7** 034015

[139] Wei Y *et al* 2016 Crystal structure manipulation of the exchange bias in an antiferromagnetic film *Sci. Rep.* **6** 28397

[140] Nozaki T, Oida M, Ashida T and Shimomura N 2014 Positive exchange bias observed in Pt-inserted $Cr_2O_3$/Co exchange coupled bilayers *Appl. Phys. Lett.* **105** 202508

[141] Kosub T *et al* 2017 Purely antiferromagnetic magnetoelectric random access memory *Nature Commun.* **8** 13985

[142] Wang Y, Zhou X, Song C, Yan Y, Zhou S, Wang G, Chen C, Zeng F and Pan F 2015 Electrical control of the exchange spring in antiferromagnetic metals *Adv. Mater.* **27** 3196-201

[143] Chiba D and Ono T 2013 Control of magnetism in Co by an electric field *J. Phys. D: Appl. Phys.* **46** 213001

[144] Yamada Y, Ueno K, Fukumura T, Yuan H T, Shimotani H, Iwasa Y, Gu L, Tsukimoto S, Ikuhara Y and Kawasaki M 2011 Electrically induced ferromagnetism at room temperature in cobalt-doped titanium dioxide *Science* **332** 1065-7

[145] Diez L H, Bernandmantel A, Vila L, Warin P, Marty A, Ono S, Givord D and Ranno L 2014 Electric-field assisted depinning and nucleation of magnetic domain walls in FePt/$Al_2O_3$/liquid gate structures *Appl. Phys. Lett.* **104** 349-51

[146] Cui B *et al* 2014 Electronic phases: reversible ferromagnetic phase transition in electrode-gated manganites *Adv. Funct. Mater.* **24** 7233-40

[147] Zhang P X, Yin G F, Wang Y Y, Cui B, Pan F and Song C 2016 Electrical control of antiferromagnetic metal up to 15 nm *Sci. China Phys. Mech. Astron.* **59** 1-5

[148] Song C, Cui B, Li F, Zhou X and Pan F 2017 Recent progress in voltage control of magnetism: Materials, mechanisms, and performance *Prog. Mater. Sci.* **87** 33-82

[149] Cheng R, Xiao D and Brataas A 2016 Terahertz antiferromagnetic spin hall nano-oscillator *Phys. Rev. Lett.* **116** 207603

[150] Cheng R, Xiao J, Niu Q and Brataas A 2014 Spin pumping and spin-transfer torques in antiferromagnets *Phys. Rev. Lett.* **113** 057601

[151] Železný J, Gao H, Výborný K, Zemen J, Mašek J, Manchon A, Wunderlich J, Sinova J and Jungwirth T 2014 Relativistic Néel-order fields induced by electrical current in antiferromagnets *Phys. Rev. Lett.* **113** 157201

[152] Barthem V M T S, Colin C V, Mayaffre H, Julien M H and Givord D 2013 Revealing the properties of $Mn_2Au$ for antiferromagnetic spintronics *Nature Commun.* **4** 2892

[153] Saidl V *et al* 2017 Optical determination of the Néel vector in a CuMnAs thin-film antiferromagnet *Nature Photon.* **11** 91-6

[154] Grzybowski M J *et al* 2017 Imaging current-induced switching of antiferromagnetic domains in CuMnAs *Phys. Rev. Lett.* **118** 057701

[155] Bodnar S Y, Šmejkal L, Turek I, Jungwirth T, Gomonay O, Sinova J, Sapozhnik A A, Elmers H J, Kläui M and Jourdan M 2017 Writing and reading antiferromagnetic $Mn_2Au$: Néel spin-orbit torques and large anisotropic magnetoresistance https://arxiv.org/abs/1706.02482

[156] Meinert M, Graulich D and Matallawagner T 2017 Key role of thermal

activation in the electrical switching of antiferromagnetic $Mn_2Au$ https://arxiv.org/abs/1706.06983

[157] Mishra R, Yu J, Qiu X, Motapothula M, Venkatesan T and Yang H 2017 Anomalous current-induced spin torques in ferrimagnets near compensation *Phys. Rev. Lett.* **118** 167201

[158] Finley J and Liu L 2016 Spin-orbit-torque efficiency in compensated ferrimagnetic cobalt-terbium alloys *Phys. Rev. Lett.* **6** 054001

[159] Bi C, Almasi H, Price K, Newhouse-Illige T, Xu M, Allen S R, Fan X and Wang W 2017 Anomalous spin-orbit torque switching in synthetic antiferromagnets *Phys. Rev. B* **95** 104434

[160] Shi G Y, Wan C H, Chang Y S, Li F, Zhou X J, Zhang P X, Cai J W, Han X F, Pan F and Song C 2017 Spin-orbit torque in MgO/CoFeB/Ta/CoFeB/MgO symmetric structure with interlayer antiferromagnetic coupling *Phys. Rev. B* **95** 104435

[161] Lambert C H *et al* 2014 All-optical control of ferromagnetic thin films and nanostructures *Science* **345** 1337-40

[162] Manz S, Matsubara M, Lottermoser T, Büchi J, Iyama A, Kimura T, Meier D and Fiebig M 2016 Reversible optical switching of antiferromagnetism in $TbMnO_3$ *Nature Photon.* **10** 653-6

[163] Fiebig M, Phuc Duong N, Satoh T, Van Aken B B, Miyano K, Tomioka Y and Tokura Y 2008 Ultrafast magnetization dynamics of antiferromagnetic compounds *J. Phys. D: Appl. Phys.* **41** 164005

[164] Kirilyuk A, Kimel A V and Rasing T 2010 Ultrafast optical manipulation of magnetic order *Rev. Mod. Phys.* **82** 2731-84

[165] Satoh T, Aken B B V, Duong N P, Lottermoser T and Fiebig M 2007 Ultrafast

spin and lattice dynamics in antiferromagnetic $Cr_2O_3$ *Phys. Rev. B* **75** 155406

[166] Fiebig M, Fröhlich D and Thiele H J 1996 Determination of spin direction in the spin-flop phase of $Cr_2O_3$ *Phys. Rev. B* **54** R12681

[167] Ivanov B A 2014 Spin dynamics of antiferromagnets under action of femtosecond laser pulses *Low Temp. Phys.* **40** 91-105

[168] Phuc Duong N, Satoh T, Fiebig M 2004 Ultrafast manipulation of antiferromagnetism of NiO *Phys. Rev. Lett.* **93** 117402

[169] Fiebig M, Fröhlich D, Lottermoser T, Pavlov V V, Pisarev R V and Weber H J 2001 Second harmonic generation in the centrosymmetric antiferromagnet NiO *Phys. Rev. Lett.* **87** 137202

[170] Sänger I, Pavlov V V, Bayer M and Fiebig M 2006 Distribution of antiferromagnetic spin and twin domains in NiO *Phys. Rev. B* **74** 144401

[171] Kimel A V, Stanciu C D, Usachev P A, Pisarev R V, Gridnev V N, Kirilyuk A and Rasing T 2006 Optical excitation of antiferromagnetic resonance in $TmFeO_3$ *Phys. Rev. B* **74** 060403

[172] Mikhaylovskiy R V, Hendry E, Kruglyak V V, Pisarev R V, Rasing T and Kimel A V 2014 Terahertz emission spectroscopy of laser-induced spin dynamics in $TmFeO_3$ and $ErFeO_3$ orthoferrites *Phys. Rev. B* **90** 184405

[173] Miyano K, Tanaka T, Tomioka Y and Tokura Y 1997 Photoinduced insulator-to-metal transition in a perovskite manganite *Phys. Rev. Lett.* **78** 4257-60

[174] Fiebig M, Miyano K, Tomioka Y and Tokura Y 1998 Visualization of the local insulator-metal transition in $Pr_{0.7}Ca_{0.3}MnO_3$ *Science* **280** 1925-8

[175] Takubo N, Ogimoto Y, Nakamura M, Tamaru H, Izumi M and Miyano K 2005 Persistent and reversible all-optical phase control in a manganite thin film

*Phys. Rev. Lett.* **95** 017404

[176] Satoh T, Kikuchi Y, Miyano K, Pollert E, Hejtmánek J and Jirák Z 2002 Irreversible photoinduced insulator-metal transition in the Na-doped manganite $Pr_{0.75}Na_{0.25}MnO_3$ *Phys. Rev. B* **65** 125103

[177] Afanasiev D, Ivanov B A, Kirilyuk A, Rasing T, Pisarev R V and Kimel A V 2016 Control of the ultrafast photoinduced magnetization across the Morin transition in $DyFeO_3$ *Phys. Rev. Lett.* **116** 097401

[178] Kimel A V, Ivanov B A, Pisarev R V, Usachev P A, Kirilyuk A and Rasing T 2009 Inertia-driven spin switching in antiferromagnets *Nature Phys.* **5** 727-31

[179] Bhattacharjee S, Bergman A, Taroni A, Hellsvik J, Sanyal B and Eriksson O 2012 Theoretical analysis of inertia-like switching in magnets: applications to a synthetic antiferromagnet *Phys. Rev. X* **2** 96-106

[180] Kim T H, Grünenberg P, Han S H and Cho B K 2017 Field-driven dynamics and time-resolved measurement of Dzyaloshinskii-Moriya torque in canted antiferromagnet $YFeO_3$ *Sci. Rep.* **7** 4515

[181] Ferguson B and Zhang X C 2002 Materials for terahertz science and technology *Nature Mater.* **1** 26-33

[182] Kampfrath T, Sell A, Klatt G, Pashkin A, Mährlein S, Dekorsy T, Wolf M, Fiebig M, Leitenstorfer A and Huber R 2011 Coherent terahertz control of antiferromagnetic spin waves *Nature Photon.* **5** 31-4

[183] Kampfrath T, Tanaka K and Nelson K A 2013 Resonant and nonresonant control over matter and light by intense terahertz transients *Nature Photon.* **7** 680-90

[184] Baierl S, Hohenleutner M, Kampfrath T, Zvezdin A K, Kimel A V, Huber R and Mikhaylovskiy R V 2016 Nonlinear spin control by terahertz-driven

anisotropy fields *Nature Photon.* **10** 715-8

[185] Mukai Y, Hirori H, Yamamoto T, Kageyama H and Tanaka K 2016 Nonlinear magnetization dynamics of antiferromagnetic spin resonance induced by intense terahertz magnetic field *New J. Phys.* **18** 013045

[186] Baierl S *et al* 2016 Terahertz-driven nonlinear spin response of antiferromagnetic nickel oxide *Phys. Rev. Lett* **117** 197201

[187] Pashkin A, Sell A, Kampfrath T and Huber R 2013 Electric and magnetic terahertz nonlinearities resolved on the sub-cycle scale *New J. Phys.* **15** 065003

[188] Seifert T *et al* 2017 Terahertz spin currents and inverse spin Hall effect in thin-film heterostructures containing complex magnetic compounds *spin* **7** 1740010

[189] Nishitani J, Nagashima T and Hangyo M 2013 Terahertz radiation from antiferromagnetic MnO excited by optical laser pulses *Appl. Phys. Lett.* **103** 081907

[190] Constable E, Cortie D L, Horvat J, Lewis R A, Cheng Z, Deng G, Cao S, Yuan S and Ma G 2014 Complementary terahertz absorption and inelastic neutron study of the dynamic anisotropy contribution to zone-center spin waves in a canted antiferromagnet $NdFeO_3$ *Phys. Rev. B* **90** 054413

[191] Jin Z, Mics Z, Ma G, Cheng Z, Bonn M and Turchinovich D 2013 Single-pulse terahertz coherent control of spin resonance in the canted antiferromagnet $YFeO_3$, mediated by dielectric anisotropy *Phys. Rev. B* **87** 094422

[192] Kim T H, Hamh S Y, Han J W, Kang C, Kee C S, Jung S, Park J, Tokunaga Y, Tokura Y and Lee J S 2014 Coherently controlled spin precession in canted antiferromagnetic $YFeO_3$ using terahertz magnetic field *Appl. Phys. Express* **7**

093007

[193] Zhou R, Jin Z, Li G, Ma G, Cheng Z and Wang X 2012 Terahertz magnetic field induced coherent spin precession in $YFeO_3$ *Appl. Phys. Lett.* **100** 061102

[194] Kubacka T *et al* 2014 Large-amplitude spin dynamics driven by a THz pulse in resonance with an electromagnon *Science* **343** 1333-6

[195] Bowlan P, Trugman S A, Bowlan J, Zhu J X, Hur N J, Taylor A J, Yarotski D A and Prasankumar R P 2016 Probing ultrafast spin dynamics in the antiferromagnetic multiferroic $HoMnO_3$ through a magnon resonance *Phys. Rev. B* **94** 100404

[196] Satoh T, Iida R, Higuchi T, Fiebig M and Shimura T 2014 Writing and reading of an arbitrary optical polarization state in an antiferromagnet *Nature Photon.* **9** 25-9

[197] Zakharov, A I, Kadomtseva, A M, Levitin, R Z and Ponyatovskii, E G 1964 Magnetic and magnetoelastic properties of a metamagnetic iron–rhodium alloy *Sov. Phys. JETP* **19**, 1348–53

**Table 1.** Summary of different magnetic manipulation methods of antiferromagnets. $T_N$ and $T_T$ represent the Néel temperature and antiferromagnetic–ferromagnetic transition temperature, respectively.

| System | Mechanism | Target | Temperature | | Ref. |
|---|---|---|---|---|---|
| NiO(111) single crystal | Strong magenetic field | Antiferromagnetic moments | $T_N$ | 551±16 K | [43] |
| $Mn_2Au$ thin films | Strong magenetic field | Antiferromagnetic domains | $T_N$ | ~1500 K | [44] |
| NiFe/IrMn/MgO/Pt stack | Exchange bias | Antiferromagnetic moments | | - | [13] |
| [Pt/Co]/IrMn-based junctions | Exchange bias | Antiferromagnetic moments | | - | [14] |
| Py/IrMn/MgO/Ta stacks | Exchange bias | Antiferromagnetic moments | | - | [65] |
| Fe/CuMnAs bilayers | Exchange bias | Antiferromagnetic spin-axis | | - | [85] |
| Co/NiO(001) bilayers | Exchange bias | Antiferromagnetic domain wall | | - | [96] |
| IrMn/NiFe Stacks | Exchange bias | Antiferromagnetic moments | | - | [98] |
| $La_{2/3}Sr_{1/3}MnO_3$/$Sr_2IrO_4$ | Exchange bias | Antiferromagnet spin-axis | $T_N$ | <240 K | [102] |
| Disordered-IrMn$_3$/insulating-$Y_3Fe_5O_{12}$ | Exchange bias | Antiferromagnetic moments | | - | [104] |
| $Fe_2CrSi/Ru_2MnGe$ bilayers | Exchange bias | Antiferromagnetic spins | | - | [106] |
| [Co/Pt]/FeMn | Exchange bias | Antiferromagnetic spins | | - | [107] |
| LSMO/SMO/LSMO | Exchange bias | Antiferromagnetic moments | | - | [108] |
| Ta/MgO/IrMn tunneling junctions | Field cooling | Antiferromagnetic spin configurations | $T_N$ | 173 K | [109] |
| FeRh/MgO | Field cooling | Antiferromagnetic moments | $T_T$ | ~400 K | [11] |

Table 2. Summary of different strain-related manipulation methods of antiferromagnets. $T_N$, $T_B$, and $T_T$ represent Néel temperature, blocking temperature, and antiferromagnetic–ferromagnetic transition temperature, respectively.

| System | Mechanism | Target | Temperature | | Ref. |
|---|---|---|---|---|---|
| $Mn_2Au$ | Magnetic anisotropy effect | Antiferromagnetic moments | $T_N$ | ~1500 K | [44] |
| IrMn/[Co/Pt] | Magnetic anisotropy effect | Antiferromagnetic moments | $T_B$ | 200 K (8 nm); | [62] |
| $Mn_2Au$ | Magnetic anisotropy effect | Antiferromagnetic domains | $T_N$ | ~1500 K | [110] |
| FeRh thin film | Metamagnetic transition | Fe spins | $T_T$ | ~350 K | [121] |
| $FeRh/BaTiO_3$ | Metamagnetic transition | Antiferromagnetic to ferromagnetic order | $T_T$ | ~350 K | [15, 123] |

Table 3. Summary of different electrical manipulation methods of antiferromagnets. $T_C$, $T_N$, $T_T$, and $T_B$ represent ferroelectric Curie temperature, Néel temperature, antiferromagnetic–ferromagnetic transition temperature, and blocking temperature, respectively.

| System | Mechanism | Target | Temperature | | Ref. |
|---|---|---|---|---|---|
| $BiFeO_3$ films | Electric field | Antiferromagnetic domains | $T_C$<br>$T_N$ | ~1100 K;<br>~640 K | [131] |
| $BiFeO_3$ bulk and films | Electric field | Spin flop | $T_C$<br>$T_N$ | ~820 $^oC$<br>~370 $^oC$ | [133] |
| Ni/NiO | Electric field | Antiferromagnetic moments | | - | [134] |
| $Cr_2O_3$ (0001) | Electric field | Antiferromagnetic domains | | - | [137] |
| $Cr_2O_3$ | Electric field | Antiferromagnetic order parameter | | - | [141] |
| [Co/Pt]/IrMn | Electric field | Antiferromagnetic spins | | - | [142] |
| [Co/Pt]/FeMn | Electric field | Antiferromagnetic moments | $T_B$ | >150 K (5 nm);<br><200 K (6 nm);<br>>200 K (15 nm) | [147] |
| $Mn_2Au$ | Electric current | Antiferromagnetic moments | $T_N$ | >1500 K | [151, 155, 156] |
| CuMnAs | Electric | Antiferromagnetic | $T_N$ | ~500 K | [154] |

| films | current | domains | | |
|---|---|---|---|---|

Table 4. Summary of different optical manipulation methods of antiferromagnets. $T_R$, $T_N$, $T_T$ and $T_G$ represent reorientation temperature, Néel temperature, antiferromagnetic–ferromagnetic transition temperature, and glassy state temperature, respectively.

| System | Mechanism | Target | Temperature | | Ref. |
|---|---|---|---|---|---|
| $TmFeO_3$ | Magnetic anisotropy effect | Antiferromagnetic spins | $T_R$ | 80–91 K | [7] |
| $TbMnO_3$ | Thermal & electronic excitation | Antiferromagnetic domains | | - | [162] |
| $Cr_2O_3$ | Spin-lattice interaction | Ultrafast spin dynamics | $T_N$ | 307.6 K | [165] |
| NiO | Magnetic anisotropy effect | $Ni^{2+}$ spins | $T_N$ | 523 K | [168] |
| $TmFeO_3$ | (Non)thermal mechanisms | Antiferromagnetic resonance | | - | [171] |
| $TmFeO_3$ and $ErFeO_3$ | Phase transition | Antiferromagnetic spins | | - | [172] |
| $Pr_{0.7}Ca_{0.3}MnO_3$ | Photocarrier injection | Antiferromagnetic to ferromagnetic order | $T_N$ | 130 K | [173, 174] |
| $Pr_{1-x}(Ca_{1-y}Sr_y)_xMnO_3$ | Light and heat interrelation | Antiferromagnetic to ferromagnetic order | $T_T$ | 110 K or 90 K | [175] |
| $Pr_{0.7}Na_{0.3}MnO_3$ | Photoexcitation | Antiferromagnetic to ferromagnetic order | $T_G$ | <45 K | [176] |
| $DyFeO_3$ | Electric field of light | Collinear to noncollinear antiferromagnetic state | | - | [177] |
| $HoFeO_3$ | Inertia-drive switching | Antiferromagnetic spins | | - | [178] |
| $YFeO_3$ | Inertia-drive switching | Antiferromagnetic spins | | - | [180] |
| NiO | Terahertz magnetic field | Antiferromagnetic spin waves | | - | [182] |
| $TmFeO_3$ | Terahertz electric field | Antiferromagnetic coherent spin oscillations | | - | [184] |
| $HoFeO_3$ crystal | Terahertz magnetic field | Nonlinear spin resonance | | - | [185] |
| NiO | terahertz magnetic field | Nonlinear spin resonance | | - | [186] |
| MnO | Photoexcitation | Antiferromagnetic magnons | $T_N$ | ~120 K | [189] |
| $NdFeO_3$ | In-plane anisotropy | Antiferromagnetic spin waves | $T_R$ | 110-170 K | [190] |
| $YFeO_3$ | Terahertz magnetic field | Antiferromagnetic spin waves and precession | $T_N$ | 645 K | [191-193] |
| $TbMnO_3$ | Magnetoelectric coupling | Antiferromagnetic spin motion | | - | [194] |
| $HoMnO_3$ | Spin-lattice thermalization | Antiferromagnetic spin order | $T_N$ | ~78 K | [195] |
| $YMnO_3$ | Optical polarization | Antiferromagnetic spin waves | | - | [196] |

## Figure Captions

**Figure 1.** Schematic of different manipulation methods of antiferromagnetic states.

**Figure 2.** Schematic of the spin configuration and exchange bias of an ferromagnet(FM)/antiferromagnet (AFM) bilayer at different stages (a)-(e) of an exchange biased hysteresis loop [47]. Copyright 1999 Elsevier Ltd.

**Figure 3.** (a) 130% magnetoresistance signal on a tunnelling device fabricated in the depicted multilayer structure with the NiFe/IrMn(1.5 nm)/MgO/Pt tunnel junction. The insets illustrate the rotation of antiferromagnetic moments in IrMn through the exchange-spring effect of the adjacent NiFe ferromagnet. (b) Hysteretic magnetoresistance of the NiFe/IrMn(1.5 nm)/MgO/Pt tunnel-junction device. (c) Field-cooled magnetization loops measured on the same wafer containing the NiFe/IrMn(1.5 nm)/MgO/Pt tunnel junction [13]. Copyright 2011 Nature Publishing Group.

**Figure 4.** (a) Magnetization loops of the stack structure with vertical *H* (⊥, squares) and parallel *H* (//, circles). Magnetoresistance acquired by sweeping *H*, which are (b) vertical to the films and (c), (d) parallel to the films, respectively. Insets of (b): schematic of no rotation (right) and exchange spring (left) of IrMn spins associated with Co/Pt magnetization. The only difference between (c) and (d) is that the in-plane *H* are orthogonal (mode 1) and parallel (mode 2) to the easy direction of IrMn, respectively, as sketched in (e) and (f) [14]. Copyright 2012 American Physical Society.

**Figure 5.** Schematic illustration of the antiferromagnetic FeRh/MgO structure and the memory writing and reading set-up. (a) Schematic of FeRh memory. For writing, the sample is cooled in a field $H_{FC}$. Black arrows denote the orientation of the magnetic moments in the ferromagnetic phase, whereas red and blue arrows denote two distinct configurations of the magnetic moments in the antiferromagnetic phase. $j$ is the readout current. (b) Resistance, R, measured at room temperature and zero magnetic field after field-cooling the sample with field parallel (blue) and perpendicular (red) to the current direction. (c) Stability of the two memory states after field-cooling (blue dot and red square) at room temperature [11]. Copyright 2014 Nature Publishing Group.

**Figure 6.** (a) Energy per FeRh unit formula versus the isotropic and anisotropic strain for antiferromagnetic (AFM) and ferromagnetic (FM) bulk FeRh. (b) Energy difference between the ferromagnetic and antiferromagnetic phases for the isotropic and anisotropic strain. The data for the transition $T^*$ temperature dependence versus isotropic strain are taken from ref. [197]. The vertical scales for both axes are chosen to emphasize the correspondence between the theory and the data [15]. Copyright 2014 Nature Publishing Group.

**Figure 7.** Schematic diagram of (001)-oriented $BiFeO_3$ crystal structure and the ferroelectric polarization (bold arrows) and antiferromagnetic plane (shaded planes). (a) Polarization with an up out-of-plane component before electrical poling. (b) 180° polarization switching mechanism with the out-of-plane component switched down by an external electrical field. 109°(c) and 71°(d) polarization switching mechanisms, with the out-of-plane component switched down by an external electrical field. PEEM

images of the same area of a $BiFeO_3$ film before (e) and after (f) electrical poling. In-plane PFM images before (g) and after (h) electrical poling [131]. Copyright 2006 Nature Publishing Group.

**Figure 8.** (a) The schematic cross-section view along the channel to the gate electrode with positive $V_G$, and the charge distribution under the effect of electric double layer. The insets are the schematic of spins in IrMn exchange spring, and an illustration of mechanical spring which enables the transfer of the force from top to the bottom. (b) $V_G$-dependent Hall resistance $R_{Hall}$ acquired by sweeping vertical $H$ [142]. Copyright 2015 Wiley.

**Figure 9.** (a) CuMnAs crystal structure and AFM ordering. (b) Optical microscopy image of the device and schematic of the measurement geometry. (c) Change in the transverse resistance after applying three successive writing pulses alternately. The reading current $J_{read}$ is applied and transverse resistance signals $R_{\perp}$ are recorded 10s after each writing pulse. A constant offset is subtracted from $R_{\perp}$ [6]. Copyright 2016 American Association for the Advancement of Science.

**Figure 10.** (a) Non-inertial and inertial models to transport a point mass over a potential [178]. Copyright 2009 Nature Publishing Group. (b) Schematic of femtosecond terahertz spin resonance. An intense free-space terahertz (THz) transient interacts with the electron spins of a sample to launch a coherent magnon wave. A femtosecond near-infrared (NIR) probe pulse co-propagating samples the induced net magnetization by means of the Faraday effect, after a variable delay time $t$ [182].

Copyright 2011 Nature Publishing Group. (c) Schematic diagram of reversible two-colour switching. Steps (I) to (IV) constitute a cycle with reversible two-color switching of a region of thickness $d_{th}(\omega_2)$ [162]. Copyright 2016 Nature Publishing Group.

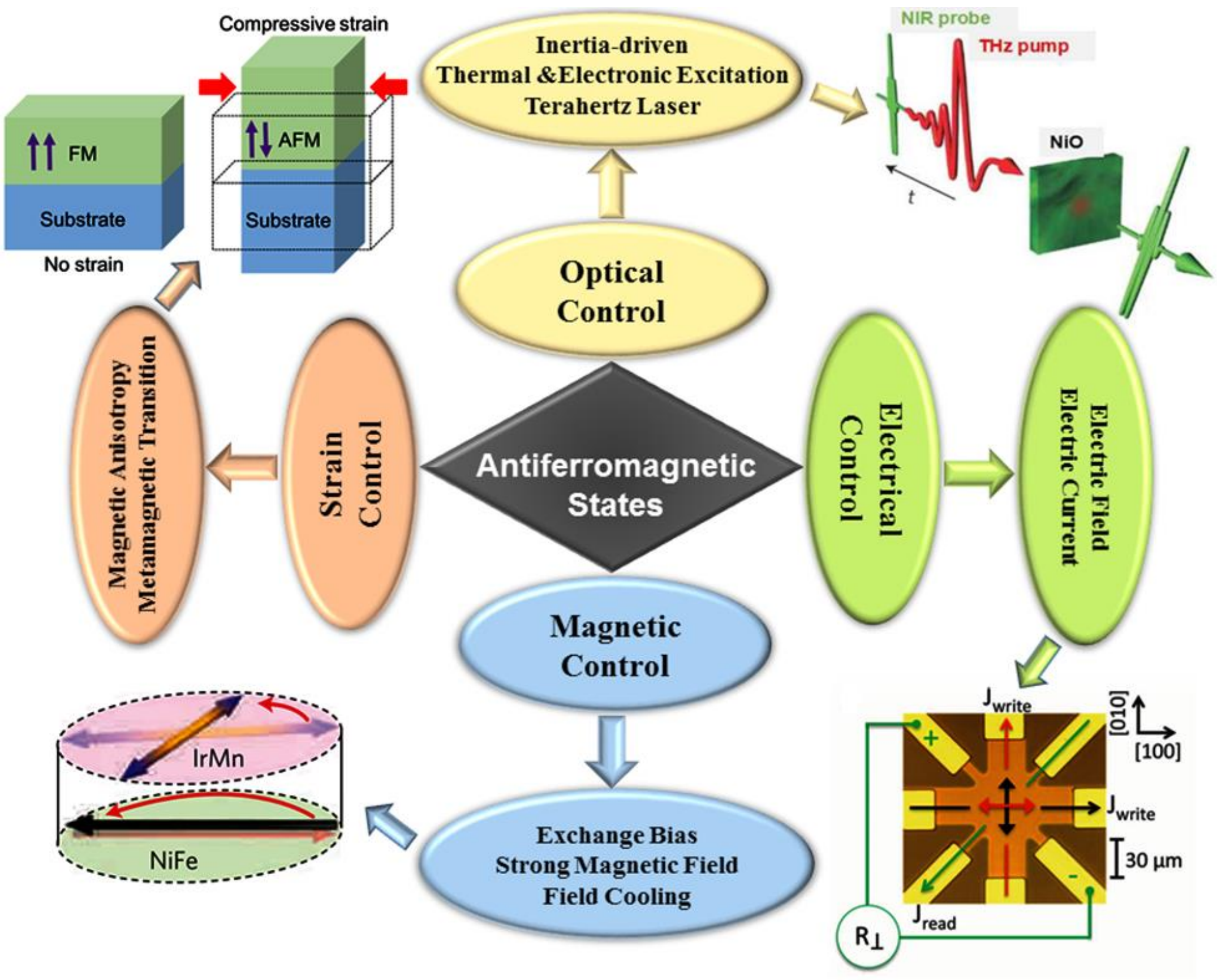

**Figure 1**

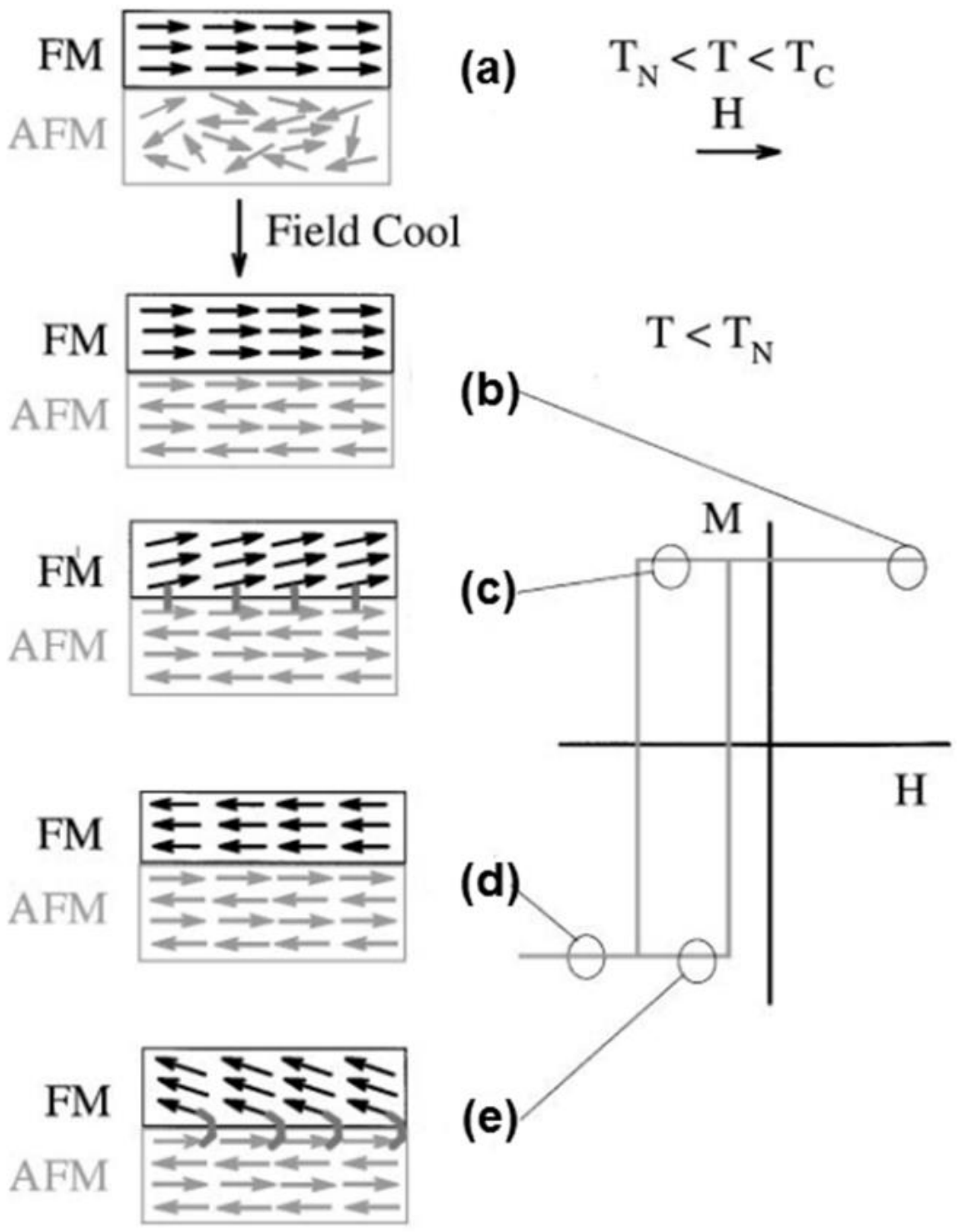

**Figure 2**

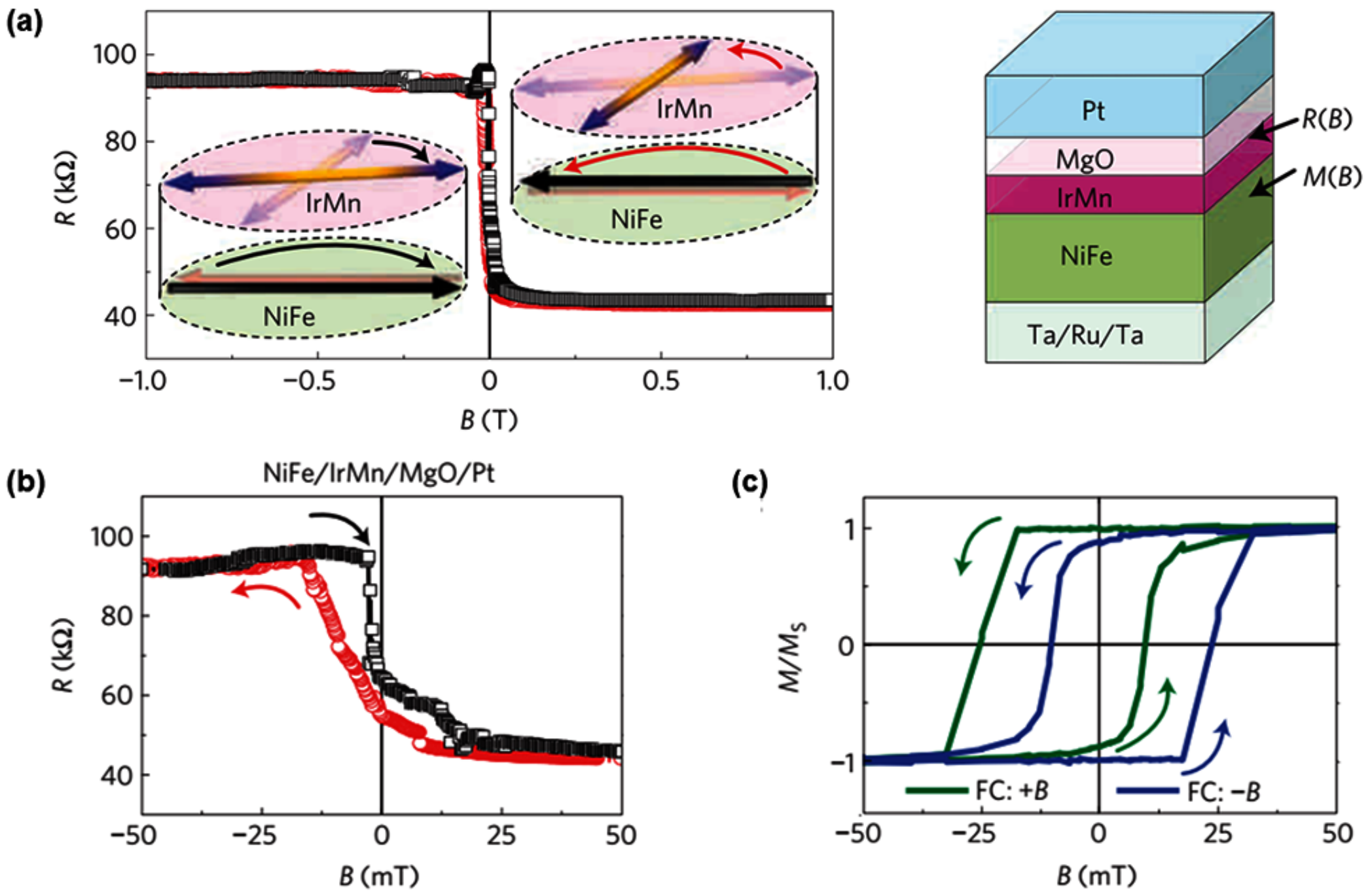

**Figure 3**

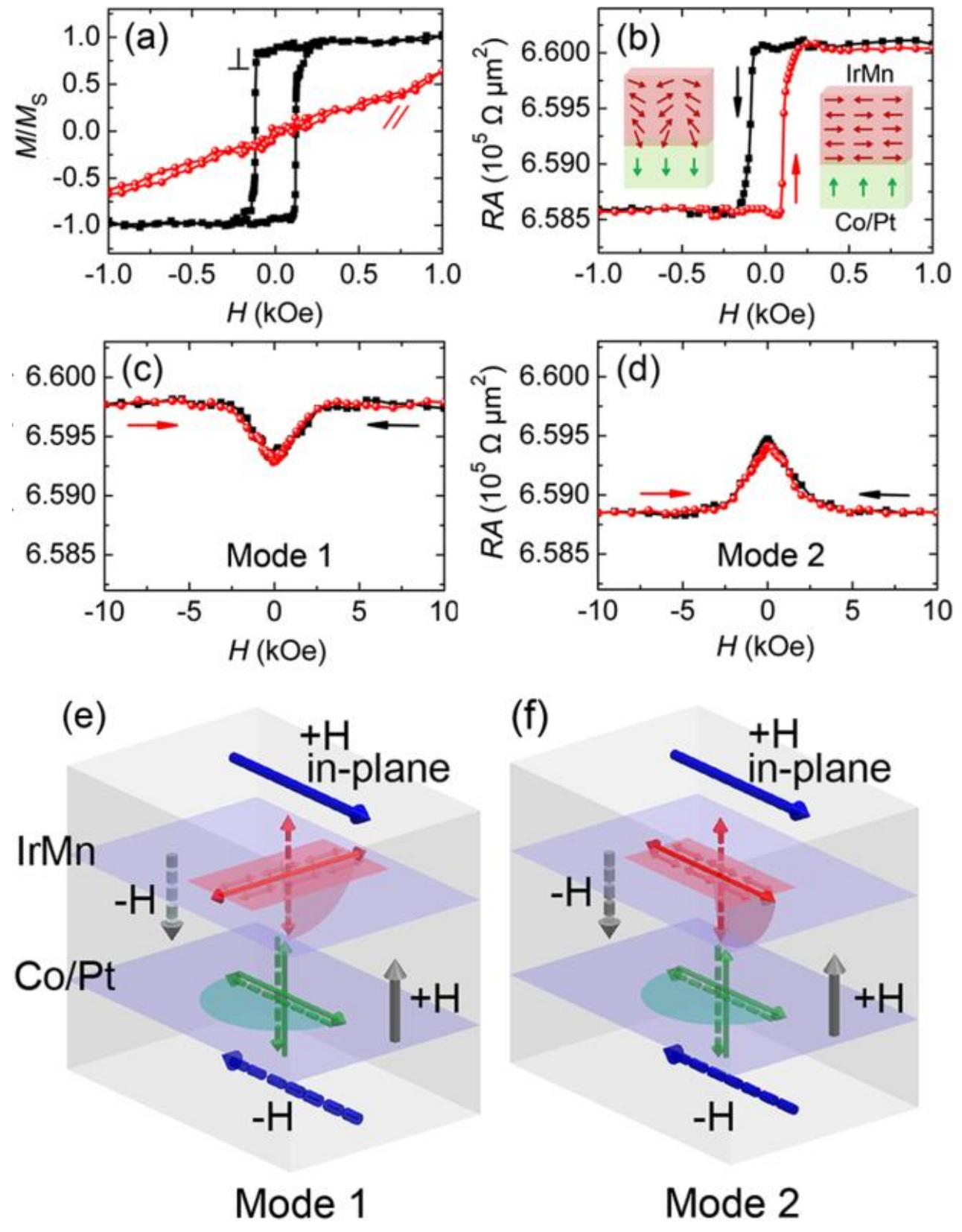

**Figure 4**

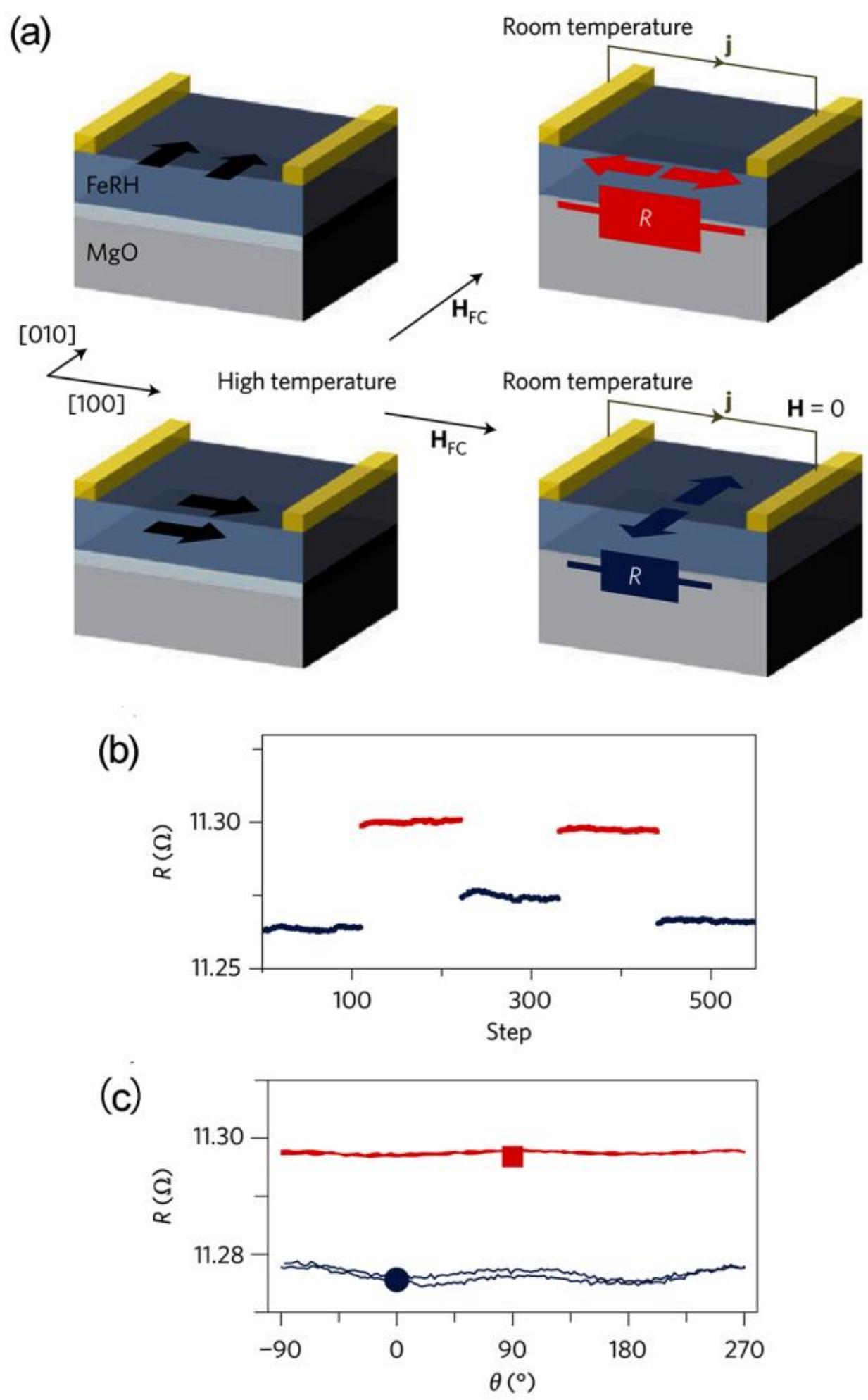

**Figure 5**

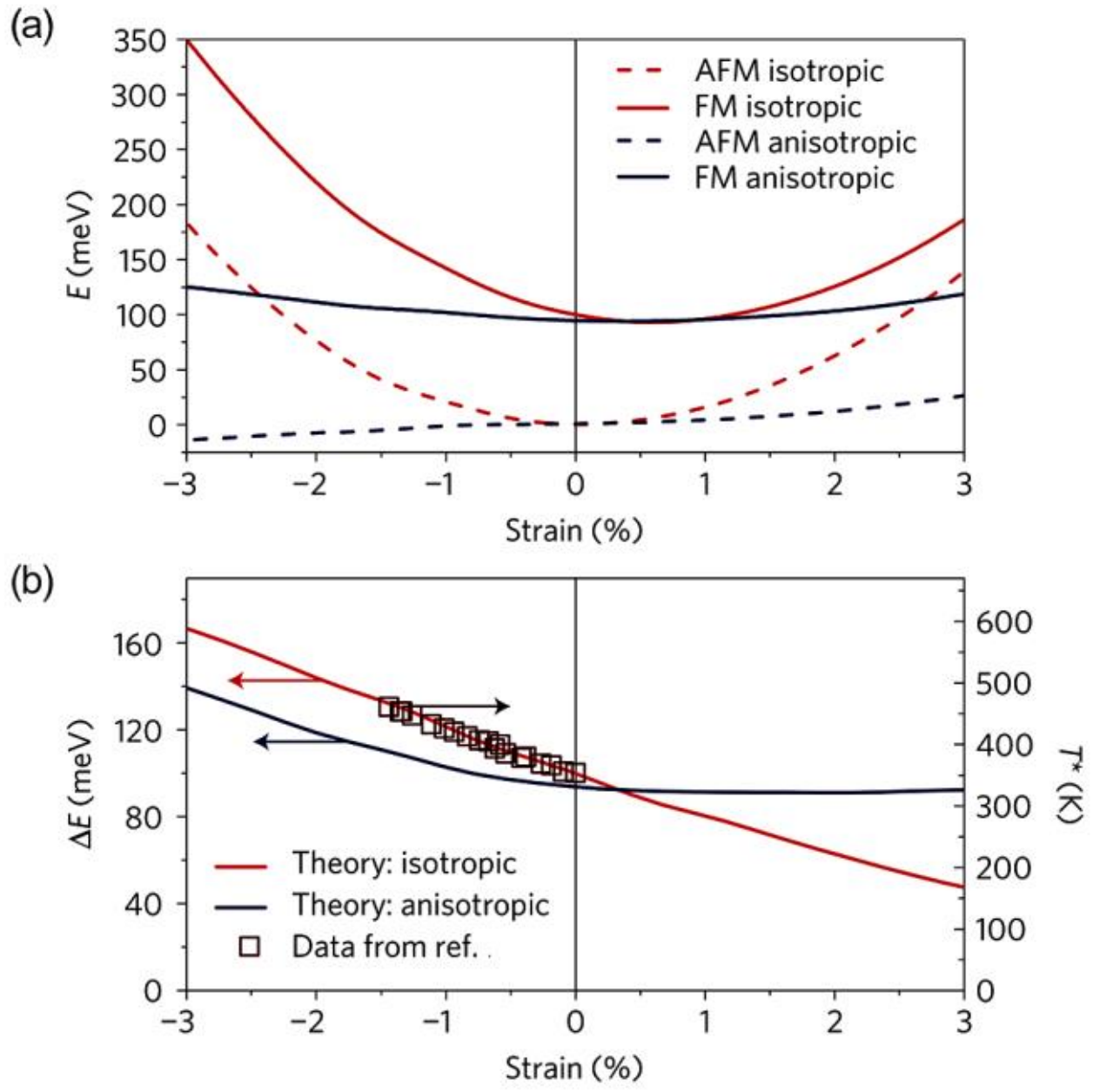

**Figure 6**

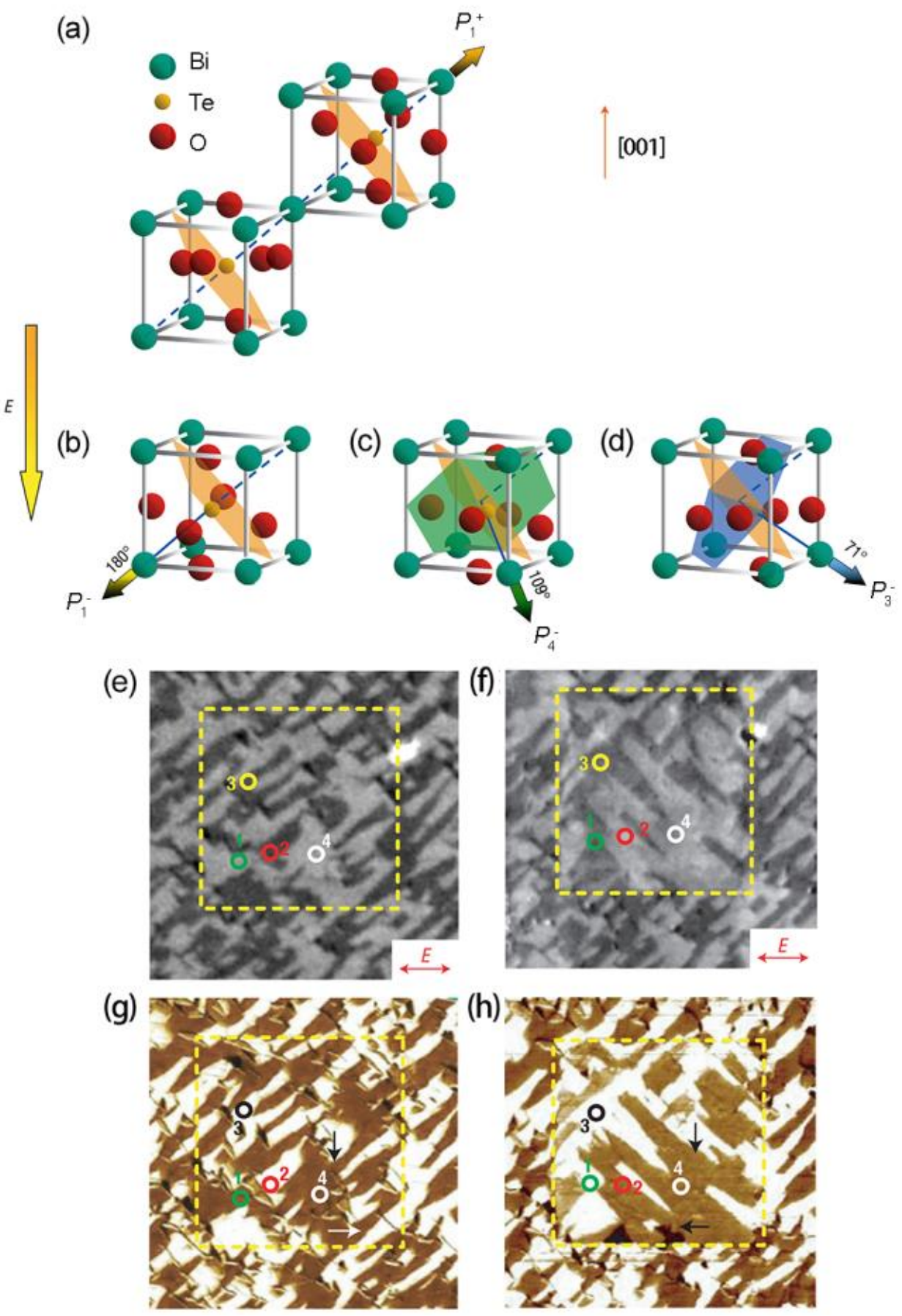

**Figure 7**

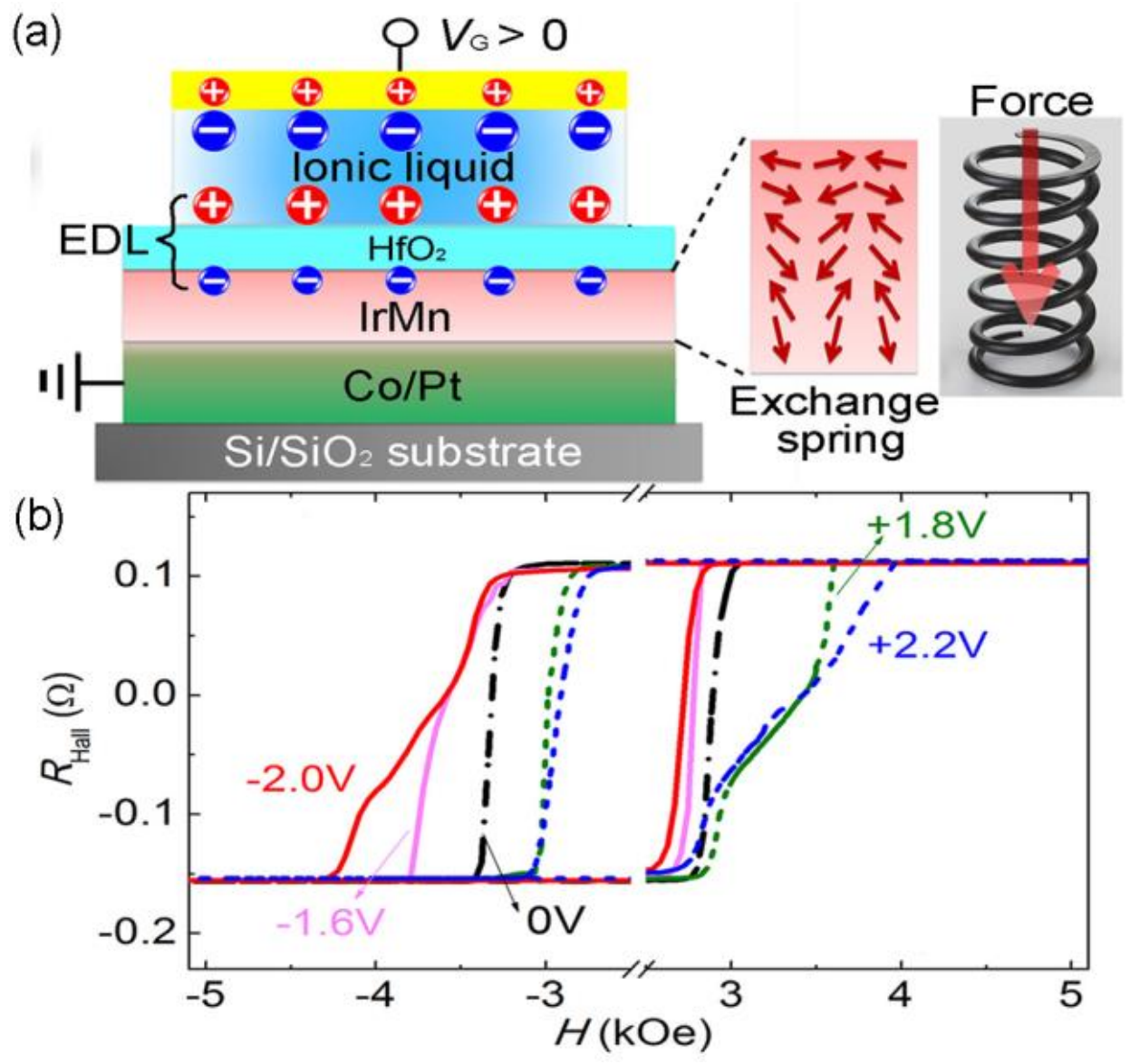

**Figure 8**

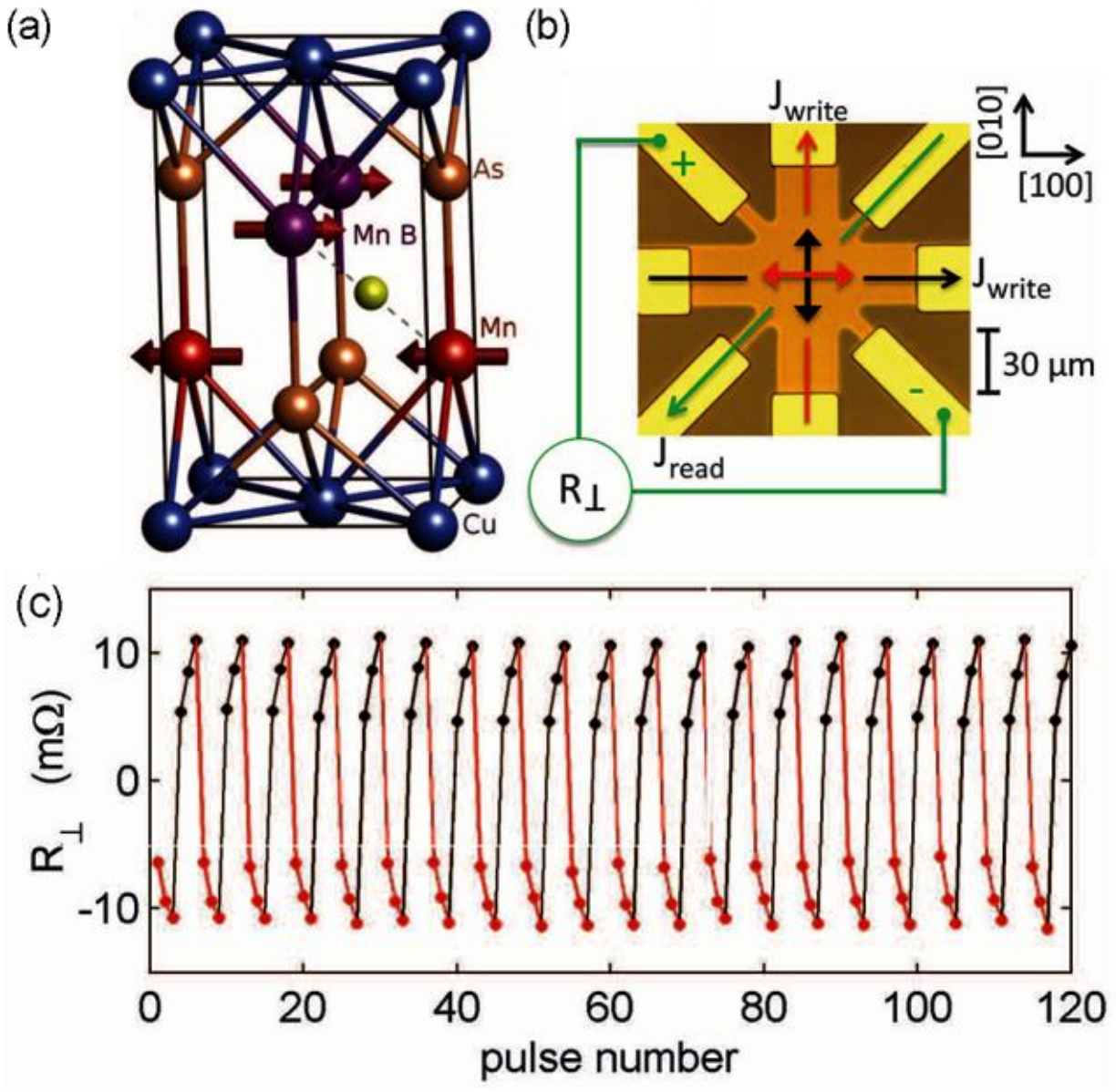

**Figure 9**

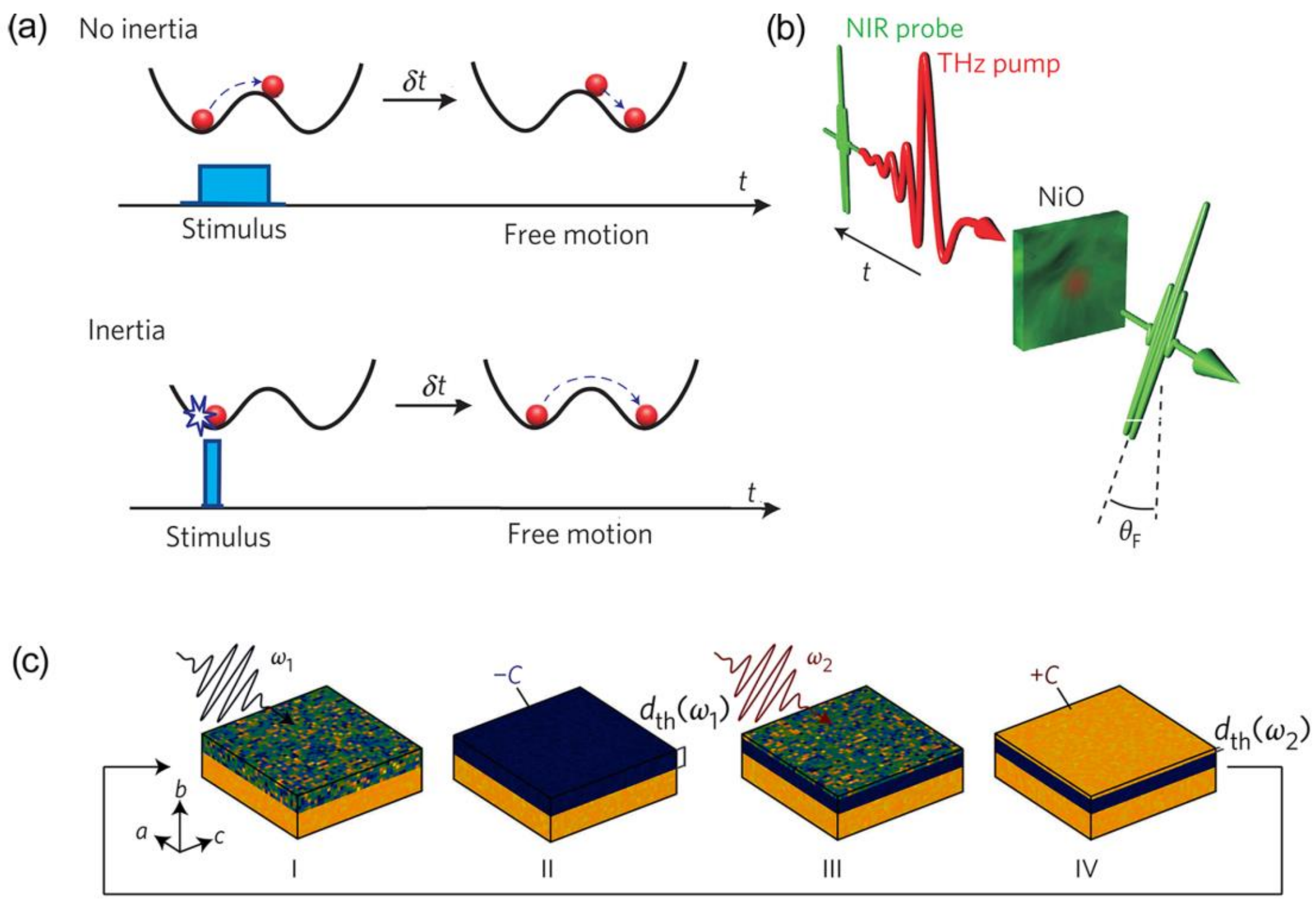

**Figure 10**